\documentclass[fleqn,usenatbib]{mnras}

\usepackage{newtxtext,newtxmath}
\usepackage[utf8]{inputenc}
\usepackage{gensymb}
\usepackage{placeins}
\usepackage{graphicx}
\usepackage{colordvi}
\usepackage{verbatim}
\usepackage{caption}
\usepackage{textcase}
\usepackage{physics}
\usepackage{natbib}
\usepackage{siunitx}
\usepackage{wrapfig}
\usepackage{subfigure}
\usepackage{amsmath}
\usepackage{tabularx}
\usepackage{hyperref}
\usepackage{filecontents}
\usepackage{makecell}
\usepackage{threeparttable}
\usepackage[normalem]{ulem}
\hypersetup{colorlinks=True, linkcolor=blue, citecolor=blue, urlcolor=blue}
\usepackage[dvipsnames]{xcolor}
\usepackage[T1]{fontenc}
\DeclareRobustCommand{\VAN}[3]{#2}
\let\VANthebibliography\thebibliography
\def\thebibliography{\DeclareRobustCommand{\VAN}[3]{##3}\VANthebibliography}

\title[SN~2018pq]{Progenitor Insights of Type IIP SN~2018pq: A Comprehensive Photometric and Spectroscopic Study}

\author[Dubey et al.]{Monalisa Dubey,$^{1,2}$\thanks{E-mail: monalisa@aries.res.in, monalisadubeyprl@gmail.com}
Kuntal Misra,$^{1}$
Raya Dastidar,$^{3,4}$
Géza Csörnyei,$^{5}$
Naveen Dukiya,$^{1,2}$
Bhavya Ailawadhi,$^{1,6}$
\newauthor
Iair Arcavi,$^{7}$
K. Azalee Bostroem,$^{8}$
Daichi Hiramatsu,$^{9,10}$
Griffin Hossienzadeh,$^{11}$
D. Andrew Howell,$^{12,13}$
\newauthor
Curtis McCully,$^{12,13}$
and Ajay Kumar Singh$^{14}$
\\
$^{1}$Aryabhatta Research Institute of Observational Sciences, Nainital-263001, India\\
$^{2}$Mahatma Jyotiba Phule Rohilkhand University, Bareilly-243006, India\\
$^{3}$Millennium Institute of Astrophysics, Nuncio Monsenor Sótero Sanz 100, Providencia, Santiago, 8320000 Chile\\
$^{4}$Instituto de Astrofísica, Universidad Andres Bello, Fernandez Concha 700, Las Condes, Santiago RM, Chile\\
$^{5}$European Southern Observatory, Karl-Schwarzschild-Straße 2, 85748 Garching, Germany\\
$^{6}$ Department of Physics, Deen Dayal Upadhyaya Gorakhpur University, Gorakhpur-273009, India\\
$^{7}$ The School of Physics and Astronomy, Tel Aviv University, Tel Aviv 69978, Israel\\
$^{8}$ Steward Observatory, University of Arizona, 933 North Cherry Avenue, Tucson, AZ 85721-0065, USA\\
$^{9}$ Center for Astrophysics \textbar{} Harvard \& Smithsonian, 60 Garden Street, Cambridge, MA 02138-1516, USA\\
$^{10}$ The NSF AI Institute for Artificial Intelligence and Fundamental Interactions, USA\\
$^{11}$ Department of Astronomy \& Astrophysics, University of California, San Diego, 9500 Gilman Drive, MC 0424, La Jolla, CA 92093-0424, USA\\
$^{12}$ Department of Physics, University of California, Santa Barbara, CA 93106-9530, USA\\
$^{13}$ Las Cumbres Observatory, 6740 Cortona Drive, Suite 102, Goleta, CA 93117-5575, USA\\
$^{14}$ Department of Applied Physics/Physics, Bareilly College, Mahatma Jyotiba Phule Rohilkhand University, Bareilly, Uttar Pradesh-243001, India
}

\date{Accepted XXX. Received YYY; in original form ZZZ}

\begin{document}
\label{firstpage}
\pagerange{\pageref{firstpage}--\pageref{lastpage}}
\maketitle
\begin{abstract}
We present high-cadence photometric and low-resolution (R $\sim$ 400--700) optical spectroscopic observations of Type IIP supernova, SN~2018pq, which exploded on the outskirts of the galaxy IC~3896A. The optically thick phase (``plateau'') lasts approximately 97 d, the plateau duration of normal Type IIP supernovae. SN~2018pq has a {\em V}-band absolute magnitude of $-16.42 \pm 0.01$ mag at 50 d, resembles normal-luminous supernova, and the {\em V}-band decline rate of 0.42$\pm$0.06 mag 50 d$^{-1}$ during the plateau phase. A steeper decline rate of 11.87$\pm$1.68 mag 100 d$^{-1}$ was observed compared to that of typical Type IIP supernovae during the transition between plateau to nebular phase. We employ detailed radiative transfer spectra modelling, \textsc{TARDIS}, to reveal the photospheric temperature and velocity at two spectral epochs. The well-fitted model spectra indicate SN~2018pq is a spectroscopically normal Type IIP supernova. Semi-analytical light curve modelling suggests the progenitor as a red supergiant star with an ejecta mass of $\sim$11 $M_\odot$ and an initial radius of 424 $R_\odot$. On the contrary, hydrodynamical modelling suggests a higher mass progenitor between 14--16 $M_\odot$.

\end{abstract}

\begin{keywords}
techniques: photometric -- techniques: spectroscopic -- transients:  supernovae -- galaxies: general -- methods: analytical -- methods: numerical
\end{keywords}



\section{Introduction}
\label{introduction}
Core-collapse supernovae (CCSNe) result from the gravitational collapse of the iron core of massive stars (M $\ge$ 8$M_\odot$; \citealt{Smartt_IIP_2009}). Type II SNe, a subclass of CCSNe, are characterised by prominent hydrogen (H) signatures in the spectra. Based on the light curve morphology, Type II are further classified into two classes: Type IIP (`plateau'), showing a prominent plateau in the light curve, and Type IIL (`linear'), displaying a linearly declining light curve \citep{Barbon_1979, Nomoto_1995}. Although recent studies have suggested a continuum in light curve shapes of Type II SNe, filling the gap between the traditional Type IIP and Type IIL subclasses \citep{Anderson_2014,Sanders_2015,Valenti_2016}. Within a volume-limited sample, Type IIP SNe are the most common, accounting for approximately 60\% of all Type II CCSNe \citep{2011MNRAS.412.1441L}. This prevalence provides a valuable opportunity to study correlations among observed photometric properties and spectroscopic parameters \citep{2003ApJ...582..905H, Anderson_2014, Valenti_2016, Gutierrez_2014, Gutierrez_2017}. 

The distinctive photometric feature of Type IIP SNe is the constant luminosity phase (plateau phase), which is attributed to the recombination of H in shock-heated expanding SN-ejecta. Type IIP SNe exhibit a diverse range of plateau durations (60--140 d after peak brightness), absolute {\em V}-band magnitudes ($-$18 $\ge$ $M_V$ $\ge$ $-$14 mag), plateau phase decline rates (0--3 mag/50d) \citep{Anderson_2014, Valenti_2016}, and ejecta velocities (1500--9600 km s$^{-1}$; \citealt{Gutierrez_2017}). The heterogeneity in light curve morphology arises from the diversity in the progenitor properties, such as mass \citep{Smartt_IIP_2009, Muller_2016, Sukhbold_2016, 2018ApJS..234...34P, Morozova_2018}, initial radius \citep{kasen_woosley_2009, Dessart_2013}, metallicity \citep{Dessart_2014, Taddia_2016}, progenitor internal structure and compositions \citep{Laplace_2021, Sukhbold_2023}, rotational speed \citep{Heger_2000, Paxton_2013}, pre-SN mass loss \citep{Smith_2014, Fuller_2020}, amount of H envelope retained before the explosion, density and structure of the immediate surrounding circumstellar medium \citep{Graham_2021, Andrews_2022, Moriya_2023}, explosion energy \citep{kasen_woosley_2009, You_2024}, opacity \citep{Kozyreva_2020, Potashov_2021}, mixing of $^{56}$Ni \citep{2018ApJS..234...34P, Kozyreva_2019}, and gamma-ray leakage \citep{Dessart_2021, Jerkstrand_2022}. After the plateau, the light curve falls exponentially, powered by the radioactive decay of synthesised $^{56}$Ni (0.001--0.360 M$_\odot$; \citealp{Anderson_2019}; \citealp{Rodriguez_Nimass}). The spectral evolution of Type IIP SNe typically features a strong P-Cygni profile of H, indicating the presence of an extended H envelope on the progenitor star before the explosion. Unlike Type IIP SNe, Type IIL SNe lack a plateau phase, and their mean apparent brightness is $\sim$ 1.5 mag brighter than that of Type IIP \citep{Patat_1994}. \cite{Faran_2014} defines Type IIL as a subgroup of Type II SNe whose {\em V}-band light curve declines by more than 0.5 mag in the first 50 d of evolution and has a weaker absorption dip of H lines than Type IIP. A continuous distribution in terms of light curve parameters, with no distinct boundary separating the two classes has been observed \cite{Anderson_2014}.

Pre-explosion imaging suggests red supergiant (RSG) stars \citep[8 to 17 $M_\odot$;][]{2009ARA&A..47...63S, Smartt_2015, Arcavi_2017} as possible progenitors of Type IIP CCSNe. Theoretical models suggest a much higher limit for the initial mass of the progenitor for all Type II SNe (e.g.  $<$40 M$_\odot$; \citealt{Fryer_1999}, $<$30 M$_\odot$; \citealt{Heger_2003}, $<$23 M$_\odot$; \citealt{ott_connor_2011},  22--25 M$_\odot$; \citealt{Muller_2016}, $\leq$27 M$_\odot$; \citealt{Sukhbold_2018}). The lack of observational detection of such massive stars is known as the ``red supergiant problem'' \citep{2009ARA&A..47...63S, Davies_2020, Kochanek_2020}. But, some studies suggest that the statistical evidence for the missing high-mass progenitors does not exist \citep{Healy_2024, Beasor_2025}. Therefore, without direct imaging of pre-SN progenitor, light curve and spectral modelling can address such unresolved questions and the characteristics of progenitor stars that significantly influence SN evolution.

For an order of magnitude estimation of the properties of the progenitor star, analytical modelling is used \citep{1980ApJ...237..541A, 1989ApJ...340..396A,  1993ApJ...414..712P}. Synthetic light curves are fit to the bolometric light curve of the observed SN and can constrain the initial properties of the progenitor, such as ejecta mass, explosion energy, opacity, ejecta velocity, and synthesised $^{56}$Ni mass. These models assume homologous expansion and a spherically symmetric SN ejecta. \cite{Nagy_2014} gave a semi-analytical model, assuming an SN with a dense core and an envelope with an exponential density profile. \cite{10.1093/mnras/staa1743} further modified the model by utilizing the Markov Chain Monte Carlo (MCMC) method and considering only the core part which resulted in the synthetic light curves failing to replicate the light curve at early stages (t < 20 d). Detailed hydrodynamical modelling could be more reliable, but, it is time-consuming and computationally expensive. 1D approximations like the combination of Module for Experiments in Stellar Astrophysics 
(\textsc{MESA}; \citealt{Paxton_2010}; \citeyear{2015ApJS..220...15P, 2018ApJS..234...34P, 2019ApJS..243...10P}) and \textsc{STELLA} \citep[the radiative transfer code available publicly;][]{1998ApJ...496..454B,2011ascl.soft08013B}, can be a useful tool to model the evolution of CCSNe through its complicated phases. \textsc{MESA+STELLA} can produce synthetic light curves based on the given initial parameters and allows for the estimation of the critical parameters for stellar evolution, such as the initial mass, metallicity, rotational speed, wind speed, mass loss rate, initial radius, explosion energy, velocity evolution, final mass, and the amount of synthesised $^{56}$Ni mass during the explosion.

\begin{figure}
    \centering
    \includegraphics[width=\linewidth,trim={0 0 2.0cm 2.0cm},clip]{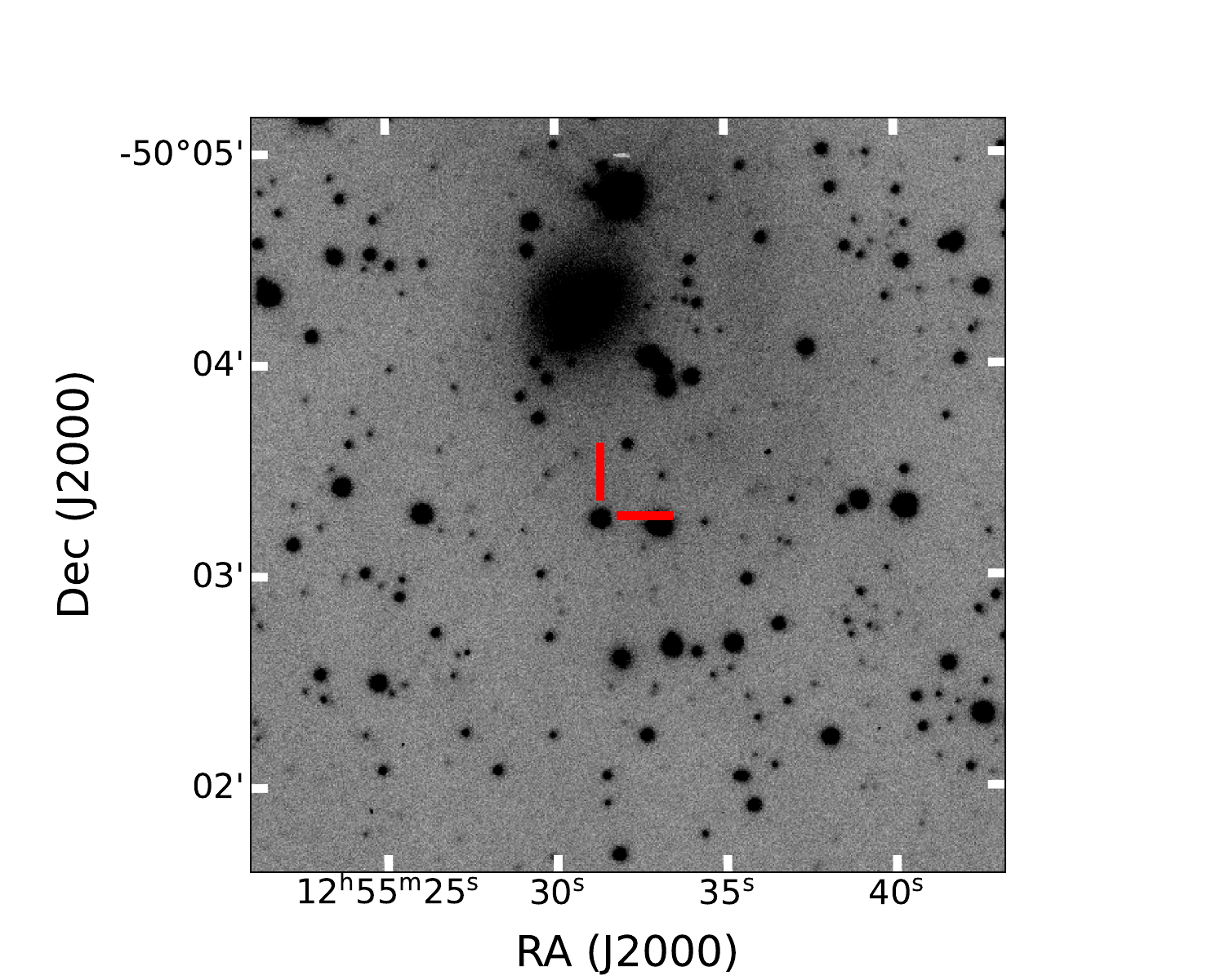}
    \caption{i-band image of SN~2018pq in IC~3896A, taken with the 1m LCO telescope on February 11$^{\rm th}$, 2018, approximately 25 d since the explosion.}
    \label{fig:SN_image}
\end{figure}

This work presents the photometric and spectroscopic analysis of a normal Type IIP SN~2018pq. The SN was discovered on February 8$^{th}$, 2018 (JD 2458157.88) at a {\em V}-band magnitude of 15.7 by the All-Sky Automated Survey for Supernovae (ASAS-SN) in the host galaxy IC~3896A \citep{2018TNSTR.177....1P}. The last non-detection was reported at {\em V}<17 mag on December 31$^{st}$, 2017. The host redshift of 0.00711$\pm$0.00003 corresponds to a distance of 23.56$\pm$1.71 Mpc\footnote{\href{https://ned.ipac.caltech.edu/byname?objname=IC+3896A&hconst=73&omegam=0.27&omegav=0.73&wmap=1&corr_z=4}{NED}} (corrected for Virgo + GA + Shapley) assuming H$_0 = 73$ km/s/Mpc, $\Omega_{\mathrm{matter}}=0.27$, and $\Omega_{\mathrm{vaccum}}=0.73$. This distance is adopted throughout the paper for further analysis. The line of sight Galactic extinction towards the host galaxy IC~3896A is $E(B-V) = 0.186$ mag \citep{milkyway_reddening}. The SN exploded on the outskirts of the host galaxy as shown in Figure~\ref{fig:SN_image}. There is little to no signature for the Na ID absorption line at the redshift of the host galaxy in the spectral sequence of SN~2018pq, indicating negligible host extinction that agrees with the SN location in the host galaxy. We, therefore, consider only galactic extinction for further analysis. The explosion epoch is estimated using GEneric cLAssification TOol \citep[GELATO;][]{2008A&A...488..383H}. The first (JD 2458160.60) and second (JD 2458162.88) spectra of SN~2018pq best match with SN~1999gi at phase 29.8 d and SN~1995ad at phase 23.9 d, respectively. The average of the explosion epochs derived from the two spectra is JD 2458135.14 ± 3.84, which is adopted as the estimated epoch of explosion, with the standard deviation taken as the associated uncertainty. The properties of SN~2018pq and its host galaxy are listed in Table~\ref{tab:SN 2018pq and Host information}.

\begin{table}
	\centering
	\caption{Basic information on SN~2018pq and its host galaxy IC~3896A. The host galaxy parameters are taken from NED.}
	\begin{tabular}{ll} 
		\hline
            \multicolumn{2}{c}{\textbf{SN~2018pq}} \\ \hline
        SN type & Type IIP\\
		RA & $12{^h}55{^m}31.290{^s}$\\
        Dec & $-$$50{^d}03{^m}16.96{^s}$ \\
	    Discovery date (JD) & 2458157.88\\
        Estimated explosion epoch (JD) & 2458135.14$\pm$3.84\\
        Distance (Mpc) & 23.56$\pm$1.71\\
        Total extinction E(B-V) (mag) & 0.186\\
	\hline\multicolumn{2}{c}{\textbf{IC 3896A}} \\ \hline
        Galaxy type & SB(rs)cd\\
        Major axis diameter (arcmin) & 3.60 \\
        Minor axis diameter (arcmin) & 3.00 \\
		Redshift & 0.00711$\pm$0.00003\\
		Helio. Velocity (km s$^{-1}$) & 1720$\pm$32\\
		\hline
	\end{tabular}
	\label{tab:SN 2018pq and Host information}

\end{table}

Section~\ref{sec:Obs&datared} of the paper outlines the observations and data reduction procedure. The light curve properties and the comparison of colour and absolute magnitude of SN~2018pq with other SNe are discussed in Section~\ref{sec:Photevo}. The identification of key spectral features and their comparison with other SNe are presented in Section~\ref{sec:spectraevol}. In the same section, radiative transfer spectral modelling using \textsc{TARDIS} is also discussed. Analytical and hydrodynamical modelling of the quasi-bolometric light curve and the derived parameters are discussed in Section~\ref{sec:modelling}. Finally, the findings of SN~2018pq based on observations and their implications are discussed in Section~\ref{sec:discussion}. 

\begin{table*}
 \begin{center}
 \caption{{\em BVgri} photometry of SN~2018pq with the LCO telescopes.}
 \label{tab:optical_phot}
 \scalebox{0.85}{
 \begin{tabular}{@{}lcccccccc}
 \hline
 Date  & JD  & Phase$^\dagger$ & {\em B} & {\em V} & {\em g} & {\em r} & {\em i} \\
 (UT) & (days) & (days) & (mag) & (mag) & (mag) & (mag) & (mag)\\ 
 \hline
		2018-02-09.31 & 2458158.81 & 23.7 & $16.38\pm0.01$  & 15.79$\pm$0.01 & 15.99$\pm$0.01 & 15.6$\pm$0.01 & 15.49$\pm$0.01 \\ 
        2018-02-11.26 &  2458160.76 & 25.6 & $16.53\pm0.01$  &  15.83$\pm$0.01   & 16.09$\pm$0.01 & 15.6$\pm$0.01 & 15.50$\pm$0.01 \\ 
        2018-02-12.10 &  2458161.60 & 26.5 &$16.51\pm0.01$ &  15.82$\pm$0.03 &  16.13$\pm$0.05 &  15.56$\pm$0.04 &  15.4$\pm$0.02 \\
        2018-02-13.23 &  2458162.73 & 27.6 &$16.66\pm0.01$ &  15.87$\pm$0.01  &  16.18$\pm$0.01&  15.59$\pm$0.01 &  15.53$\pm$0.01 \\
        2018-02-17.18  & 2458166.68 & 31.5 &$16.79\pm0.01$  & 15.91$\pm$0.01 &   16.27$\pm$0.01 & 15.61$\pm$0.01 &  15.51$\pm$0.01 \\
        2018-02-20.91  & 2458170.41 & 35.3 &$16.97\pm0.02$  & 15.98$\pm$0.02 &  16.29$\pm$0.03 &   15.70$\pm$0.01 &  15.53$\pm$0.03 \\
        2018-02-25.18  & 2458174.68  & 39.5 &$17.06\pm0.01$ &  15.96$\pm$0.01  & 16.43$\pm$0.01 &  15.66$\pm$0.01  & 15.52$\pm$0.01 \\
        2018-03-01.14  & 2458178.64  & 43.5 &$17.15\pm0.01$ &  15.99$\pm$0.01  & 16.49$\pm$0.01 &  15.66$\pm$0.01  & 15.49$\pm$0.01 \\
        2018-03-04.85  & 2458182.35  & 47.2 &$17.23\pm0.02$ &  16.02$\pm$0.01  & 16.54$\pm$0.01 &  15.67$\pm$0.01  & 15.48$\pm$0.01 \\
        2018-03-09.39  & 2458186.89  & 51.8 &$17.32\pm0.01$ &  16.02$\pm$0.01  & 16.58$\pm$0.01 &  15.69$\pm$0.01  & 15.48$\pm$0.01 \\
        2018-03-12.72 & 2458190.22  & 55.1 &$17.42\pm0.02$ &  16.05$\pm$0.01  &  16.50$\pm$0.01 &  15.71$\pm$0.01  & 15.50$\pm$0.01 \\
        2018-03-17.30  & 2458194.80  & 59.7 &$17.47\pm0.01$ &  16.07$\pm$0.01  & 16.69$\pm$0.01 &   15.70$\pm$0.01  & 15.47$\pm$0.01 \\
        2018-03-20.88  & 2458198.38  & 63.2 &$17.59\pm0.01$ &  16.09$\pm$0.01  & 16.76$\pm$0.01 &  15.71$\pm$0.01  & 15.49$\pm$0.01 \\
        2018-03-23.06  & 2458200.56  & 65.4 &$17.70\pm0.02$ &  16.13$\pm$0.01  & 16.81$\pm$0.01 &  15.73$\pm$0.01  & 15.50$\pm$0.01 \\
        2018-03-29.08  & 2458206.58  & 71.4 &$17.65\pm0.02$ &  16.15$\pm$0.01  & 16.79$\pm$0.01 &  15.74$\pm$0.01  &  15.51$\pm$0.01 \\
        2018-04-04.20  & 2458212.70  & 77.6 &$17.82\pm0.02$ &  16.25$\pm$0.01  & 16.93$\pm$0.01 &  15.81$\pm$0.01  & 15.57$\pm$0.01 \\
        2018-04-11.70  & 2458220.20  & 85.1 &$18.12\pm0.02$ &  16.43$\pm$0.01  & 17.18$\pm$0.01 &  15.97$\pm$0.01  & 15.71$\pm$0.01 \\
        2018-04-17.56  & 2458226.06  & 90.9 &$18.41\pm0.02$ &  16.66$\pm$0.01  & 17.46$\pm$0.01 &  16.17$\pm$0.01  & 15.91$\pm$0.01 \\
        2018-04-23.74 & 2458232.24  & 97.1 &$19.17\pm0.09$ &  17.33$\pm$0.02  & 18.22$\pm$0.02 &  16.72$\pm$0.01  & 16.47$\pm$0.01 \\
        2018-04-25.12  & 2458233.62  & 98.5 &$19.39\pm0.04$ &  17.71$\pm$0.01  & 18.52$\pm$0.01 &  17.04$\pm$0.01  & 16.81$\pm$0.01 \\
        2018-04-28.11  & 2458236.61  & 101.5 &$19.90\pm0.17$ &  18.41$\pm$0.04  & 19.34$\pm$0.06 &  17.54$\pm$0.02  & 17.28$\pm$0.06 \\
        2018-04-30.36  & 2458238.86  & 103.7 &$20.28\pm0.41$ &  18.47$\pm$0.03  & 19.44$\pm$0.05 &  17.71$\pm$0.01  & 17.45$\pm$0.01 \\
        2018-05-02.05  & 2458240.55  & 105.4 & --  &  18.61$\pm$0.10   & 19.51$\pm$0.16 &   17.80$\pm$0.05  & 17.54$\pm$0.05 \\
        2018-05-03.06  & 2458241.56  &  106.4 & -- &  18.73$\pm$0.08  &    -- &  17.84$\pm$0.02  & 17.51$\pm$0.03 \\
        2018-05-04.38  & 2458242.88  & 107.7 & --  & 18.79$\pm$0.02  & -- &  17.84$\pm$0.01  & 17.59$\pm$0.01 \\
        2018-05-08.26  & 2458246.76  & 111.6 &$20.69\pm0.06$ &  18.74$\pm$0.02  & 19.66$\pm$0.02 &  17.87$\pm$0.01  & 17.61$\pm$0.01 \\
        2018-05-12.22  & 2458250.72  & 115.6 &$20.61\pm0.06$ &  18.81$\pm$0.01  & 19.74$\pm$0.02 &  17.93$\pm$0.01  & 17.66$\pm$0.01 \\
        2018-05-16.24  & 2458254.74  & 119.6 &$20.58\pm0.04$ &  18.81$\pm$0.01  & 19.69$\pm$0.01 &  17.97$\pm$0.01  & 17.71$\pm$0.01 \\
        2018-05-20.38  & 2458258.88  & 123.7 &--   &  18.95$\pm$0.03  &  --   &   18.05$\pm$0.01  & 17.78$\pm$0.02 \\
        2018-05-22.22  & 2458260.72  & 125.6 &$20.57\pm0.04$ &  18.88$\pm$0.01  & 19.72$\pm$0.02 &   18.00$\pm$0.01  & 17.82$\pm$0.02 \\
        2018-05-25.00  & 2458264.50  & 129.4 &$20.72\pm0.16$ &   --   &    --   &      --  &    --    \\
        2018-05-26.02  & 2458264.52  & 129.4 &$20.62\pm0.12$ &  18.85$\pm$0.03  & 19.81$\pm$0.04 &  18.05$\pm$0.02  & 17.87$\pm$0.02 \\
        2018-06-02.06  & 2458271.56  & 136.4 &$20.76\pm0.07$ &  18.98$\pm$0.02  & 19.82$\pm$0.03 &  18.09$\pm$0.01  & 17.96$\pm$0.01 \\
        2018-06-10.71  & 2458280.21  & 145.1 &$20.73\pm0.16$ &  19.02$\pm$0.03  & 20.01$\pm$0.03 &  18.16$\pm$0.01  & 18.04$\pm$0.02 \\
        2018-06-20.09  & 2458289.59  & 154.5 &$20.63\pm0.05$ &  19.14$\pm$0.02  & 19.82$\pm$0.02 &  18.24$\pm$0.01  & 18.15$\pm$0.01 \\
        2018-06-30.11  & 2458299.61  & 164.5 &$20.74\pm0.36$ &  19.24$\pm$0.09  & 19.88$\pm$0.12 &  18.36$\pm$0.03  & 18.30$\pm$0.05 \\
        2018-07-08.09  & 2458307.59  & 172.5 &$20.80\pm0.05$ &  19.33$\pm$0.02  & 20.02$\pm$0.02 &  18.38$\pm$0.01  & 18.39$\pm$0.01 \\
        2018-07-13.67  & 2458313.17  & 178.1 &$21.06\pm0.11$ &  19.39$\pm$0.02  & 20.07$\pm$0.01 &  18.46$\pm$0.01  & 18.44$\pm$0.01 \\
        2018-07-26.97  & 2458326.47  & 191.3 &$20.79\pm0.19$ &  19.57$\pm$0.08  &  --   &     --  &     --  \\
        2018-07-27.01  & 2458326.51  & 191.4 &--    &    --   &  --    &  18.57$\pm$0.05  & 18.56$\pm$0.04 \\
        2018-07-28.43  & 2458327.93  & 192.8 &$20.70\pm0.29$ &  19.63$\pm$0.08  &  --   &    18.61$\pm$0.07  &  18.50$\pm$0.26 \\
        2018-07-30.17  & 2458329.67  & 194.5 &$20.96\pm0.06$ &  19.53$\pm$0.02  & 20.16$\pm$0.05 &  18.57$\pm$0.01  & 18.61$\pm$0.02 \\
        2018-08-12.38  & 2458342.88  & 207.7 &--   &    --   &  20.33$\pm$0.02 &  18.76$\pm$0.01  & 18.78$\pm$0.02 \\
        2018-08-13.69  & 2458344.19  & 209.1 &$21.19\pm0.05$ &   19.70$\pm$0.02  & 20.36$\pm$0.03 &  18.75$\pm$0.02  &   --    \\
        2018-08-14.00  & 2458344.50  & 209.4 & --  &       -- &    --  &   18.78$\pm$0.02  & 18.83$\pm$0.02 \\

	 \hline
 \end{tabular}}
 \end{center}
 $^\dagger$Phase with respect to the explosion epoch (JD = 2458135.14).
\end{table*}

\begin{table}
 \begin{center}
 \caption{Log of spectroscopic observations.}
 \label{Table:spec_log}
 \scalebox{1.0}{
 \begin{tabular}{@{}lccr}
 \hline
 Date & JD & Phase$^\dagger$ & Exposure time \\
 (UT) & (days) & (days) & (s)\\
 \hline
2018-02-11.60 & 2458160.60 & 25.5 & 1800.00\\
2018-02-13.38$^\ddagger$ & 2458162.88 & 27.7 & 180.00\\
2018-03-09.75 & 2458186.75 & 51.6 & 2700.00\\
2018-03-17.56 & 2458194.56 & 59.4 & 2700.00\\
2018-03-29.52 & 2458206.52 & 71.4 & 2700.00\\
2018-04-06.61 & 2458214.61 & 79.5 & 2700.00\\
 \hline
 \end{tabular}}
 \end{center}
  $^\dagger$Phase with respect to the explosion epoch (JD = 2458135.14).
  
  $^\ddagger$ NTT/EFOSC spectrum.
\end{table}

\section{Observation and data reduction} 
\label{sec:Obs&datared}

Multi-epoch imaging of SN~2018pq was acquired with the 1 m telescopes furnished with the Sinistro camera in {\em BVgri} filters of the Las Cumbres Observatory \citep[LCO;][]{Brown_LCO_2013} network of telescopes as part of the Global Supernova Project (GSP). The photometric observations started on February 9$^{th}$, 2018, one day after the discovery, and continued till August 14$^{th}$, 2018, covering the plateau, fall from the plateau, and the nebular phase of the SN. The data reduction and PSF photometry of the SN was performed using the \texttt{lcogtsnpipe}\footnote{\url{https://github.com/LCOGT/lcogtsnpipe/}} pipeline \citep{Valenti_2016}. Colour terms and nightly zero points for each filter were determined using the AAVSO Photometric All-Sky Survey \citep[APASS;][]{apass_2016} catalogue, which were later used to obtain the calibrated SN magnitudes. The multi-colour photometry of SN~2018pq is presented in Table~\ref{tab:optical_phot}. 

Low resolution (R $\sim$ 400--700) spectroscopic observations at five epochs from February 11$^{th}$ to April 6$^{th}$, 2018 were acquired with the FLOYDS spectrograph mounted on the LCO 2m telescopes using a 2 arcsec slit. The \texttt{floydsspec}\footnote{\url{https://github.com/svalenti/FLOYDS\textunderscore pipeline/}} pipeline \citep{Velenti_2014} was used to extract 1D spectra including the wavelength and flux calibration. We include the publicly available spectrum observed on February 13$^{th}$, 2018 (JD 2458162.88) acquired by the extended Public ESO Spectroscopic Survey of Transient Objects team \citep[ePESSTO+,][]{Smartt_pessto_2015} with the EFOSC spectrograph (R $\sim$ 300--600) using 1 arc-sec slit mounted on the 3.6m ESO-New Technology Telescope (NTT). All spectra were scaled by the \texttt{lightcurve-fitting}\footnote{\url{https://github.com/griffin-h/lightcurve_fitting}} module \citep{Griffin_LCfit_2022} to recover the flux loss due to slit size and corrected for the galactic redshift. The log of spectroscopic observation is presented in Table~\ref{Table:spec_log}.

\section{Light curve features}
\label{sec:Photevo}
\begin{figure}
\includegraphics[width=0.9\columnwidth ]{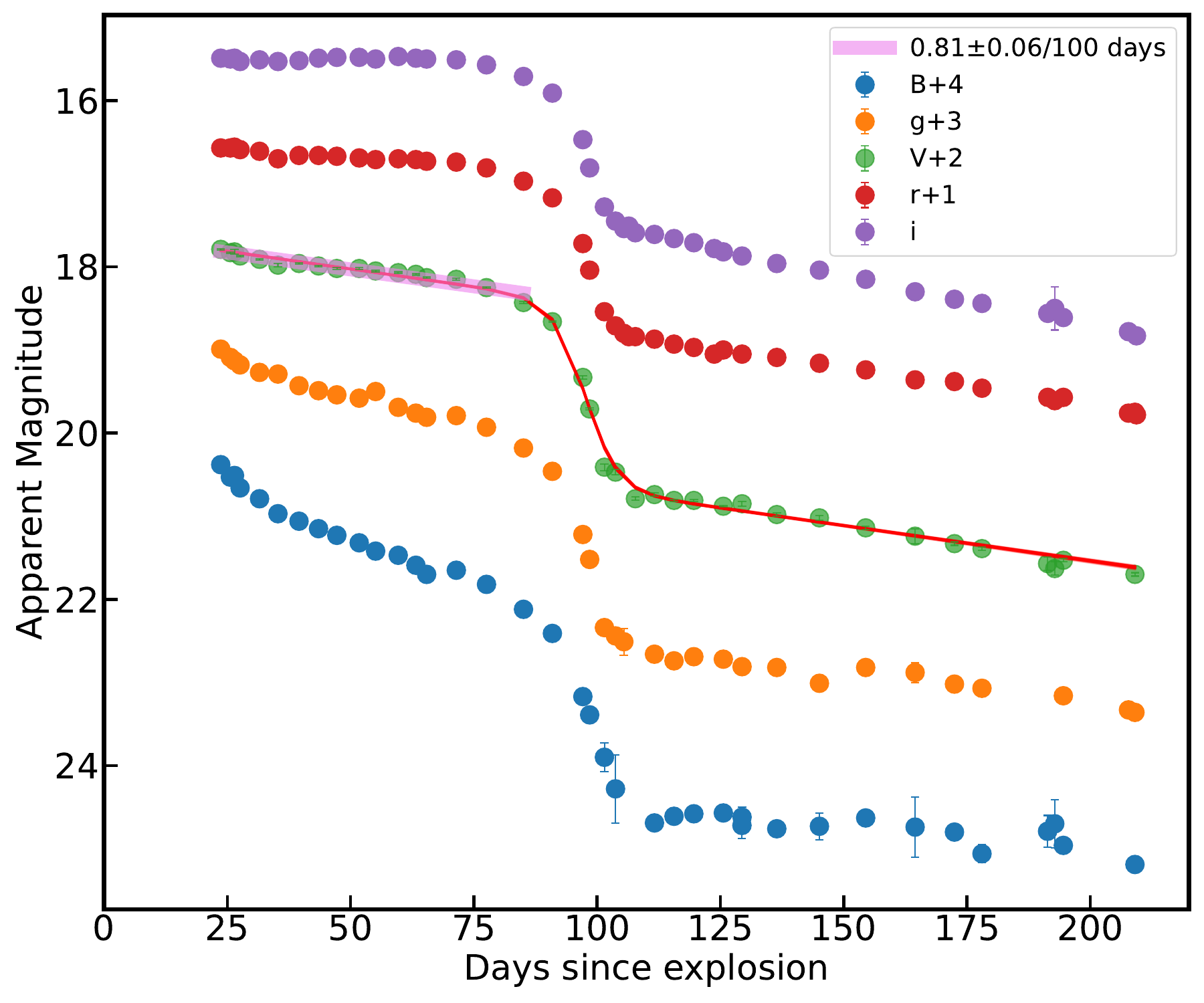} 
\caption{The {\em BVgri} light curves of SN~2018pq from $\sim$ 23 to 209 d since the explosion spanning the plateau, fall from the plateau and the nebular phase. A thick pink line in the {\em V} band light curve represents the slope during the plateau phase. The best fit of the analytic function \citep{Valenti_2016} to the {\em V}-band light curve is shown by a red solid line.}
\label{fig:valenti}
\end{figure}

The observed light curves of Type II SNe show a wide range in their properties. Studying the evolution of light curves allows us to understand their diverse origins and reveal important physical processes behind the explosions. The multi-band temporal evolution of SN~2018pq spanning from $\sim$23 d to 209 d after the explosion is shown in Figure~\ref{fig:valenti}. The early rise and peak of the light curve are missing in our observations, due to the large gap ($\sim$23 d) between the explosion and discovery of the SN. The light curve, however, has a good cadence in the plateau, the fall from the plateau to the radioactive tail, and the nebular phase. The {\em V}-band absolute magnitude at 50 d after the explosion is $-$16.42$\pm$0.01 mag with the decline rates during the plateau phase in {\em BgVri} bands are 2.6$\pm$0.14, 1.6$\pm$0.10, 0.81$\pm$0.06, 0.47$\pm$0.06 and 0.15$\pm$0.07 mag 100 d$^{-1}$, respectively.

The following analytical function \citep{Valenti_2016} is used to extract the light curve parameters. 

\begin{equation} 
y(t)= \frac{-a_0}{1+e^\frac{t-t_{PT}}{w_0}} + p_0\times t + m_0
\label{eu_Valenti}
\end{equation}

In this equation, the depth and duration of the drop from the plateau to the nebular phase are represented by $a_0$ and 6$\times$$w_0$, respectively. $t$ is the epoch of the explosion, $t_\mathit{PT}$ is the plateau duration in days, the parameter $p_0$ constrains the slope before and after the drop and $m_0$ is a constant. The first term in the above equation represents the Fermi-Dirac function, which describes the phase transition between the plateau and the nebular phase. To replicate the slope of the light curve before and after the drop, the second component adds the slope to the Fermi-Dirac function. This analytical function was fit to the {\em V}-band light curve of SN~2018pq using the MCMC method with the \texttt{emcee}\footnote{\url{https://emcee.readthedocs.io/en/stable/}} module.
The estimated best-fit parameters for $t_\mathit{PT}$, $a_0$, $w_0$, and $p_0$ are 97.57$\pm$0.05 d, 2.24$\pm$0.01 mag, 3.30$\pm$0.07 d, and 0.0084$\pm$0.0082 mag/day, respectively. The best-fit analytical function to the {\em V}-band light curve is shown in Figure~\ref{fig:valenti} with a red solid line.

To compare the light curve and spectral properties of SN~2018pq with other Type IIP SNe, we constructed a comparison sample that constitutes SNe with varying plateau lengths and luminosities. The distance, reddening and other physical parameters of the SNe in the comparison sample are listed in Table~\ref{tab:comparison_objects}.

\begin{table*}
\centering
\renewcommand{\arraystretch}{1.5} 
\begin{threeparttable}
\caption{Properties of the Type IIP SNe in the comparison sample.}
\label{tab:comparison_objects}
\begin{tabular}{ccccccccccc}
\hline \hline
Supernova & \makecell{Parent \\ Galaxy} & \makecell{Distance \\ (Mpc)} & \makecell{A$^{tot}_V$ \\ (mag)} & \makecell{M$^{50}_{V}$ \\ (mag)} & \makecell{$t_\mathit{tp}$ \\ (days)} & \makecell{E($10^{51}$) \\ (ergs)} & \makecell{R \\ ($R_\odot$)} & \makecell{M$_\mathit{ej}$ \\ ($M_\odot$)} & \makecell{$^{56}$Ni \\ ($M_\odot$)} & Ref. \\
\hline

1999em & NGC 1637 & 11.7(0.1) & 0.31 & $-$16.69$\pm$0.01 & 95 & $0.5-1$ & $120-150$ & $10-11$ & $0.042^{+0.027}_{-0.019}$ & 1, 2, 3 \\

2005cs & M51 & 7.1(1.2) & 0.16 & $-$14.83$\pm$0.10 & $\sim$130 & 0.3 & -- & 8--13 & 0.003 & 4\\

2012aw & NGC 3351 & 9.9(0.1) & 0.23  & $-$16.67$\pm$0.04 & 96$\pm$11 &0.9$\pm$0.3 & 337$\pm$67 & 14$\pm$5 & 0.06$\pm$0.01 & 5 \\

2013ab & NGC 5669 & 24.0(0.9) & 0.14 & $-$16.70$\pm$0.10 & $\sim$78d & $\sim$0.35 & $\sim$600 & $\sim$7 & 0.064$\pm$0.003 & 6 \\ 

2016gfy & NGC 2276 & 29.64(2.65) & 0.65 & $-$16.74$\pm$0.22  & $\sim$102.5 & 0.90$\pm$0.15 & 310$\pm$70 & 13.2$\pm$1.2 & 0.033$\pm$0.003 & 7 \\

2019edo & NGC 4162 & 36.18(1.83) & 0.097 & $-$16.12 & $\sim$76 & 0.8 & 500 & 6.6 & $\sim$0.05 & 8 \\

2021gmj & NGC 3310 & 17.8$^{+0.6}_{-0.4}$ & 0.153 & $-$15.45 & $\sim$100 & 0.294 & -- & 10 & 0.014$\pm$0.001 & 9, 10 \\  

2021yja &  NGC 1325 &  21.8 & 0.32 & $-$17.5 & $\sim$125 & 1.53 & 631 & 15 & 0.175-0.2 & 11, 12\\

\bfseries 2018pq & \bfseries IC 3896A & \bfseries 	23.56(1.71) & \bfseries 0.578 & \bfseries $-$16.42$\pm$0.01 & \bfseries $\mathbf{97.57^{+0.05}_{-0.06}}$ & $\mathbf{2.567^{+0.189}_{-0.687}}$ & $\mathbf{ 424.32^{+6.89}_{-256.01}}$ & $\mathbf{ 10.660^{+0.294}_{-1.778}}$ & \bfseries  $\mathbf{0.029\pm0.003}$ & \bfseries This work \\  
\hline
\end{tabular}
References: (1) \cite{Hamuy_2001}, (2) \cite{Leonard_2002}, (3) \cite{Elmhamdi_2003_1999em}, (4) \cite{2005cs_pastorello}, (5) \cite{2012aw_bose}, (6)\cite{2013ab_bose}, (7) \cite{2016gfy_singh}, (8) \cite{2019edo_Tsvetkov}, (9) \cite{2021gmj_Retamal}, (10) \cite{2021gmj_Murai}, (11) \cite{2021yja_kozyreva}, (12) \cite{2021yja_hosseinzadeh}
\end{threeparttable}
\end{table*}

\begin{figure}
\includegraphics[width=\columnwidth]{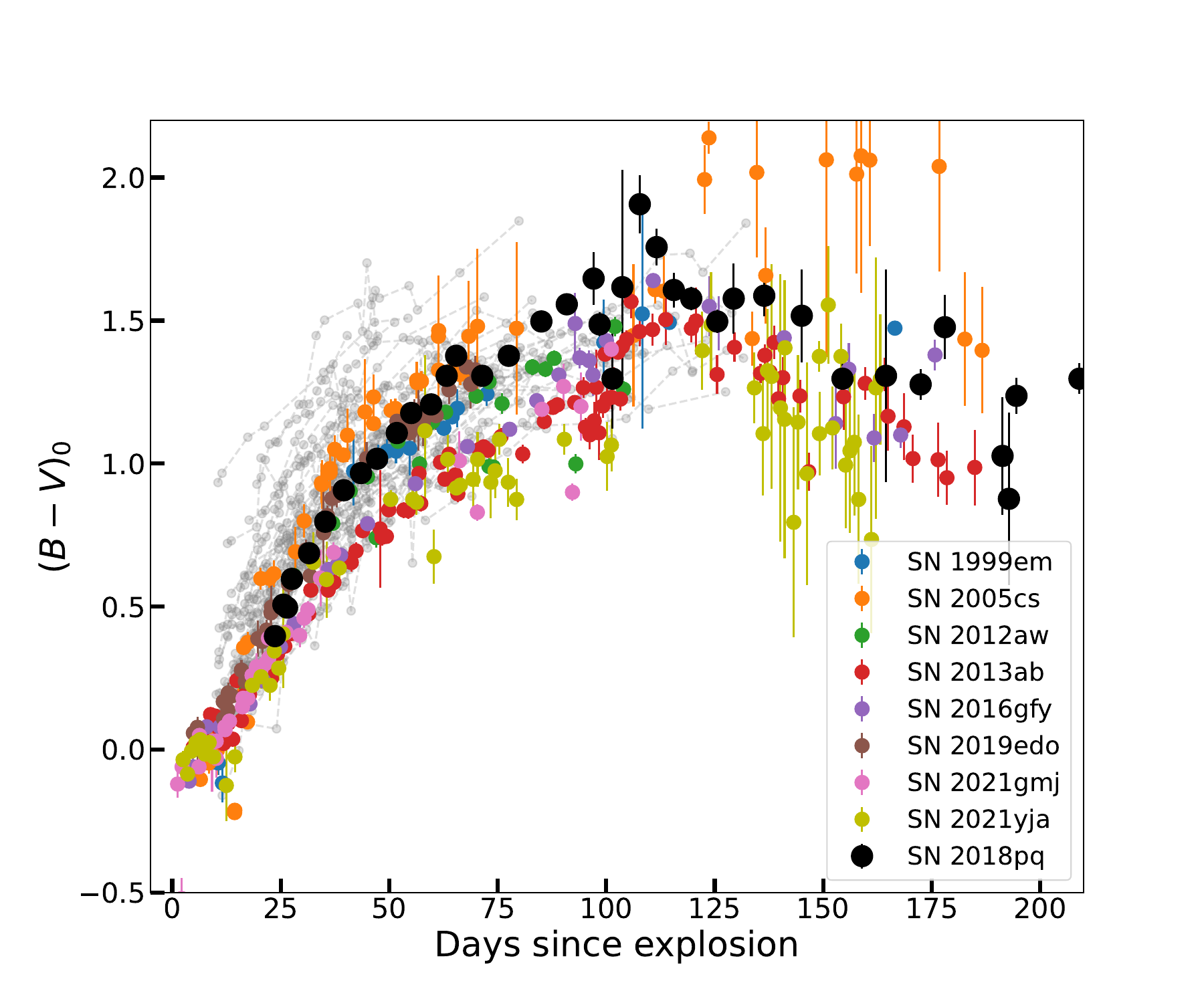}
\caption{Comparison of the {\em (B-V)$_0$} colour of SN~2018pq with other Type IIP SNe. The gray points represent the sample from \citet{deJaeger_2016esw_2018}.}
\label{fig:colour_plot}
\end{figure}

Figure~\ref{fig:colour_plot} depicts the {\em (B-V)$_0$} colour evolution of SN~2018pq, SNe of the comparison sample and the samples taken from \cite{deJaeger_2016esw_2018} represent with gray points. The colours of all SNe are computed after correcting for reddening. The temporal evolution of reddening-corrected broadband colours reveals crucial information about the dynamics of the SN ejecta. We notice that the {\em (B-V)$_0$} colour of SN~2018pq gradually becomes redder by about 1.75 mag until 111.6 d since explosion, representing the expansion and cooling of the ejecta over the first 100 d. This colour evolution is similar to SNe~2012aw and 2019edo till 75 d post-explosion. After the end of the plateau phase, the colour shifts blueward. At this stage, the radiation from radioactive decay, mainly from $^{56}$Ni, is prominent and heats the SN ejecta.

\begin{figure}
\includegraphics[width=1.0\linewidth ]{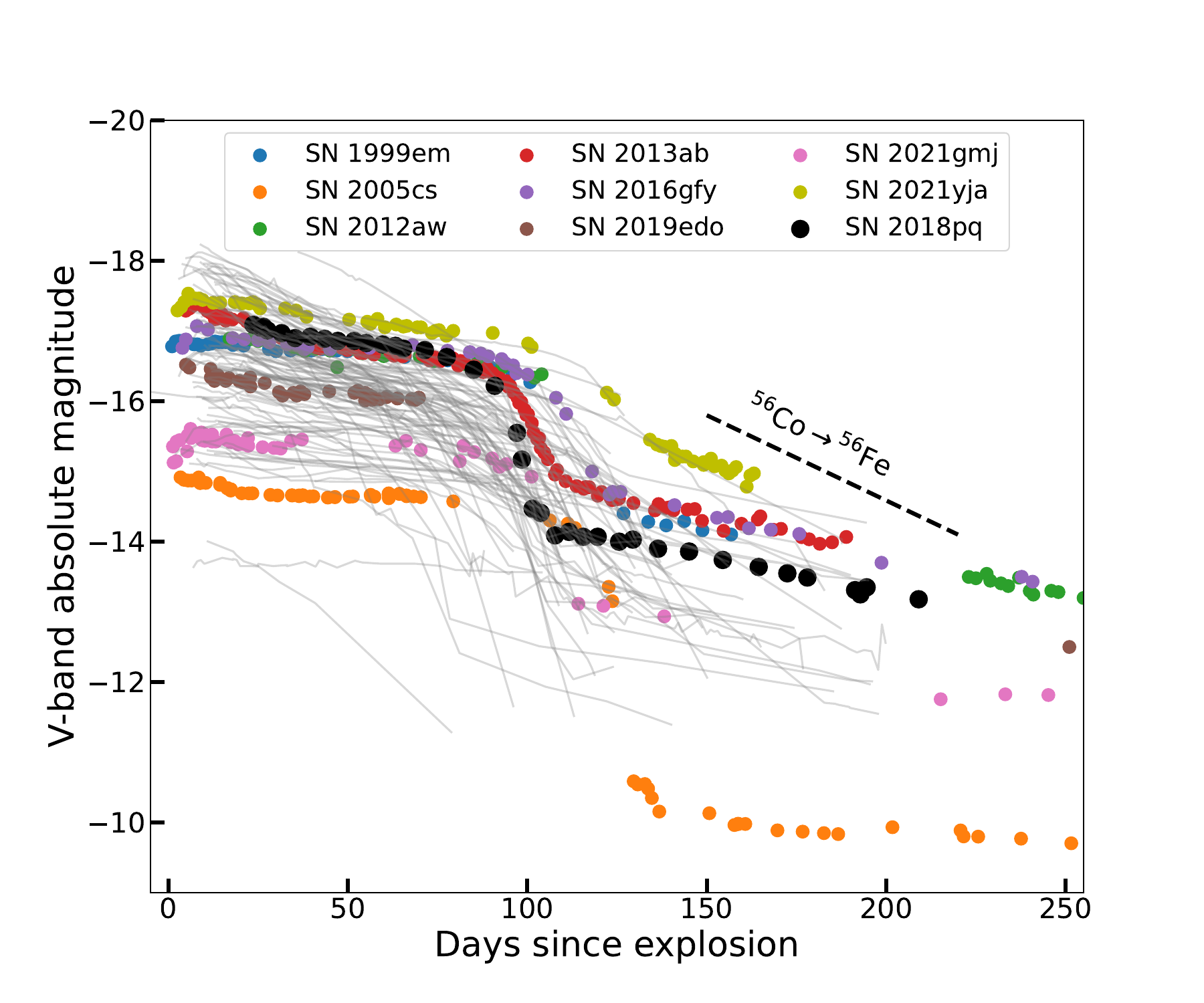} 
\caption{Comparison of {\em V}-band absolute magnitude of SN~2018pq with other Type IIP SNe. The comparison sample (gray lines) corresponds to the Type II SN sample from \citet{Anderson_2014}.}
\label{fig:abs_mag}
\end{figure}

The {\em V}-band absolute magnitude light curve of SN~2018pq, Type IIP SNe of the comparison sample and the sample SNe taken from \cite{Anderson_2014} represent in gray lines are shown in Figure~\ref{fig:abs_mag}. The absolute magnitudes of the SNe are corrected for distance and extinction values listed in Table~\ref{tab:comparison_objects}. The {\em V}-band absolute magnitude of SN~2018pq at 50 d after the explosion is $-16.42\pm0.01$, like normal Type IIP SNe~1999em and 2012aw. The decline of SN~2018pq (11.87$\pm$1.68 mag 100 d$^{-1}$) from the plateau to the nebular phase is steeper than other SNe. A similar trend was seen in SN~2013ab (8.7$\pm$0.2 mag 100 d$^{-1}$; \citealp{2013ab_bose}). The drop at the transition is 2.24$\pm$0.01 mag, similar to SN~2021gmj \citep{2021gmj_Murai, 2021gmj_Retamal}. Hence, SN~2018pq exhibits a plateau length of 97.57$\pm$0.05 d, similar to other normal Type IIP SN, such as SN~1999em (95 d) and SN~2012aw (96 d), but with a sharp decline during phase transition, which signifies a rapid decrease in luminosity. The decay rate in the nebular phase was 0.99$\pm$0.03 mag 100 d$^{-1}$, compatible with the expected decay rate of $^{56}$Co to $^{56}$Fe (0.98 mag 100 d$^{-1}$). 

\subsection{Estimation of $^{56}$Ni mass}
\label{Nimass}
The deeper layers of the SN ejecta can be probed during the optically thin nebular phase. At this time, the ejecta is transparent to optical photons but is still opaque to gamma rays, and the luminosity is proportional to the synthesised $^{56}$Ni. Using the tail luminosity, we estimated the amount of $^{56}$Ni synthesised during the explosion employing the \citet{2003ApJ...582..905H} expression (equation~\ref{eu_Mni}):

\begin{equation} \label{eu_Mni}
M_\mathit{Ni} = 7.866\times10^{-44}\times L_t \exp\left[\frac{\left(\frac{t-t_0}{1+z}\right)-6.1}{111.26}\right ] M_\odot
\end{equation}

\noindent
where $t_0$ is explosion time, 6.1 d is the half-life of $^{56}$Ni, and 111.26 d is the e-folding time of $^{56}$Co. The luminosity in the nebular phase, $L_t$ (equation~\ref{eu_Lt}; \citealt{2003ApJ...582..905H}), is estimated from the {\em V}-band magnitude corrected for distance, extinction, and bolometric correction (BC) factor of $0.26\pm0.06$ \citep{hamuy2001PhD}. 

\begin{equation} \label{eu_Lt}
log(L_t) = -0.4[V_t - A_\mathit{v}(tot)+BC]+2logD - 3.256
\end{equation}

\noindent
where D represents the distance of the SN in centimetres. We estimate $M_{Ni}$ at the last five epochs in the nebular phase. To accurately estimate the error in $M_{Ni}$, we use Bayesian inference and compare the $V_t$ values predicted by \autoref{eu_Lt} to the observed ones, taking into account the uncertainties in $V_t$, BC, distance, $t_0$, and z. We run 10000 iterations of MCMC simulations and find the value of $M_{Ni}$ is 0.025$\pm$0.004 M$_\odot$.

The $^{56}$Ni mass of SN~1987A ($0.075\pm0.005$ $M_\odot$) was accurately determined in \cite{Arnett_1996}. Assuming similar gamma-ray leakage from the SN ejecta and comparing the quasi-bolometric light curves (constructed using the {\em BVgri} bands) of both SNe~1987A and 2018pq, the synthesised $^{56}$Ni mass can be estimated using equation~\ref{eu_87A} \citep{Sapiro2014}:

\begin{equation} \label{eu_87A}
 M_\mathit{Ni} = [0.075\times (L_\mathit{SN}/L_\mathit{87A})] M_\odot
\end{equation}

\noindent
Here, $L_\mathit{87A}$ and $L_\mathit{SN}$ are the luminosities at the tail phase of the quasi-bolometric light curves of SNe~1987A and 2018pq, respectively. For SN~2018pq, we estimated the $^{56}$Ni mass using equation~\ref{eu_87A} at the same five epochs considered in the previous method, accounting for the uncertainties in $L_\mathit{SN}$, and $L_\mathit{87A}$. The weighted average of these individual estimates is considered to be the $^{56}$Ni mass. Additionally, we add in quadrature the distance uncertainty, resulting in a $^{56}$Ni mass of 0.039$\pm$0.014 M$_\odot$.

$^{56}$Ni mass can also be estimated from the slope of the drop from the plateau phase to the nebular phase. \cite{Elmhamdi_2003A&A} used the steepness parameter $S = -dM_v/dt$, defined as the slope of transition from the plateau to the radioactive tail phase, to determine the $^{56}$Ni mass. S is anti-correlated with $^{56}$Ni mass. An increase in $^{56}$Ni mass enhanced the degree of mixing and has a larger contribution to radiative diffusion at the end of the plateau, causing a lower S value. The modified relation, by \cite{Singh_2018MNRAS} with a larger sample of Type II SNe, between $^{56}$Ni mass and S is represented in equation~\ref{eu_Mnislope}:

\begin{equation} \label{eu_Mnislope}
log M(^{56}Ni) = -(3.5024\pm0.0960)\times S - 1.0167\pm0.0034
\end{equation}

\noindent
Using the steepness parameter, S = $11.87\pm1.68$ mag 100 d$^{-1}$ for SN~2018pq, the estimated $^{56}$Ni was found to be 0.037$\pm$0.006 M$_{\odot}$. 

The synthesised $^{56}$Ni mass has been determined from the above-mentioned methods. The weighted average of $^{56}$Ni mass from different methods is $0.029\pm0.003$ $M_\odot$ taken as the amount of $^{56}$Ni synthesised during the explosion. 

\section{Spectral analysis}
\label{sec:spectraevol}

\begin{figure}
\includegraphics[width=0.9\linewidth ]{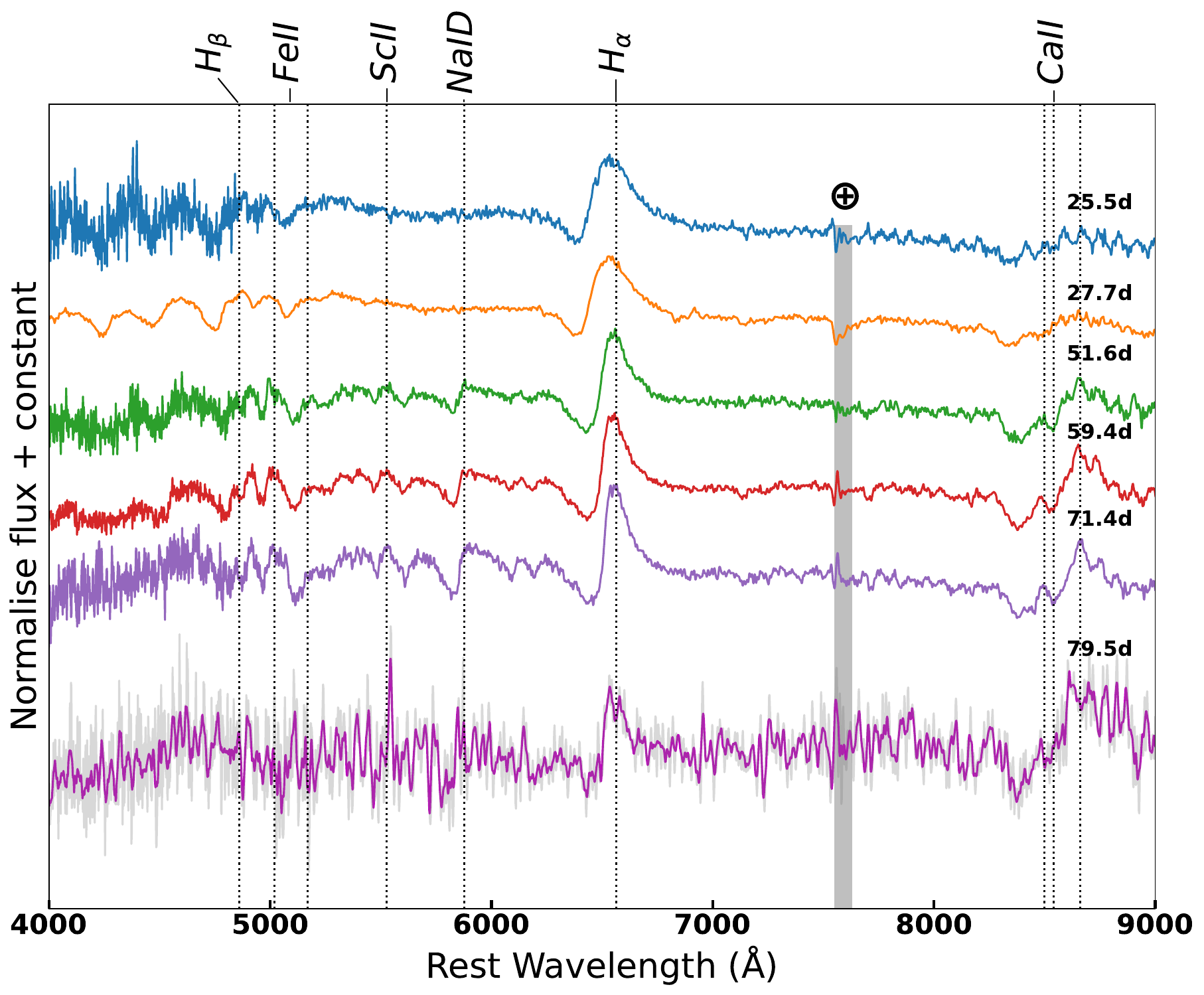} 
\caption{Spectral evolution of SN~2018pq spanning from $\sim$25 to 79.5 d since explosion shows the presence of the P-Cygni profile of $H\alpha$ along with several metal lines (such as Fe II, Sc II, Ca II NIR triplet) throughout the plateau phase.}
\label{fig:Spectra_evo}
\end{figure}

The redshift corrected spectra from $\sim$25 d to $\sim$79 d after the explosion are shown in Figure~\ref{fig:Spectra_evo}.  All spectra are normalised with their median value and prominent absorption lines are marked. Spectra display the P-Cygni profile of $H\alpha$ line, one of the main characteristic properties of Type II SNe. In the photospheric phase, strong H Balmer lines represent an extended H envelope and a gradual density profile with high expansion velocity, which is usually seen in the spectra of Type IIP SNe. Metal lines (such as Fe II, Sc II, Ca II NIR triplet) become stronger with time, revealing the formation of several heavy ions.

\subsection{Key spectroscopic features}

The first spectrum of SN~2018pq was obtained at the early photospheric phase. In a simplified setting, assuming a grey atmosphere, a photosphere is defined as a surface at which the optical depth reaches $\tau$ = 2/3, and the temperature is equal to the effective temperature \citep{Sim_2017}. In this phase, due to the high optical thickness of SN ejecta, photons cannot escape directly from the inner region; as a result, we get pseudo-continuum spectra with discrete spectral features in this phase of evolution.

\begin{figure}
    \centering
    \includegraphics[width=1.0\linewidth ]{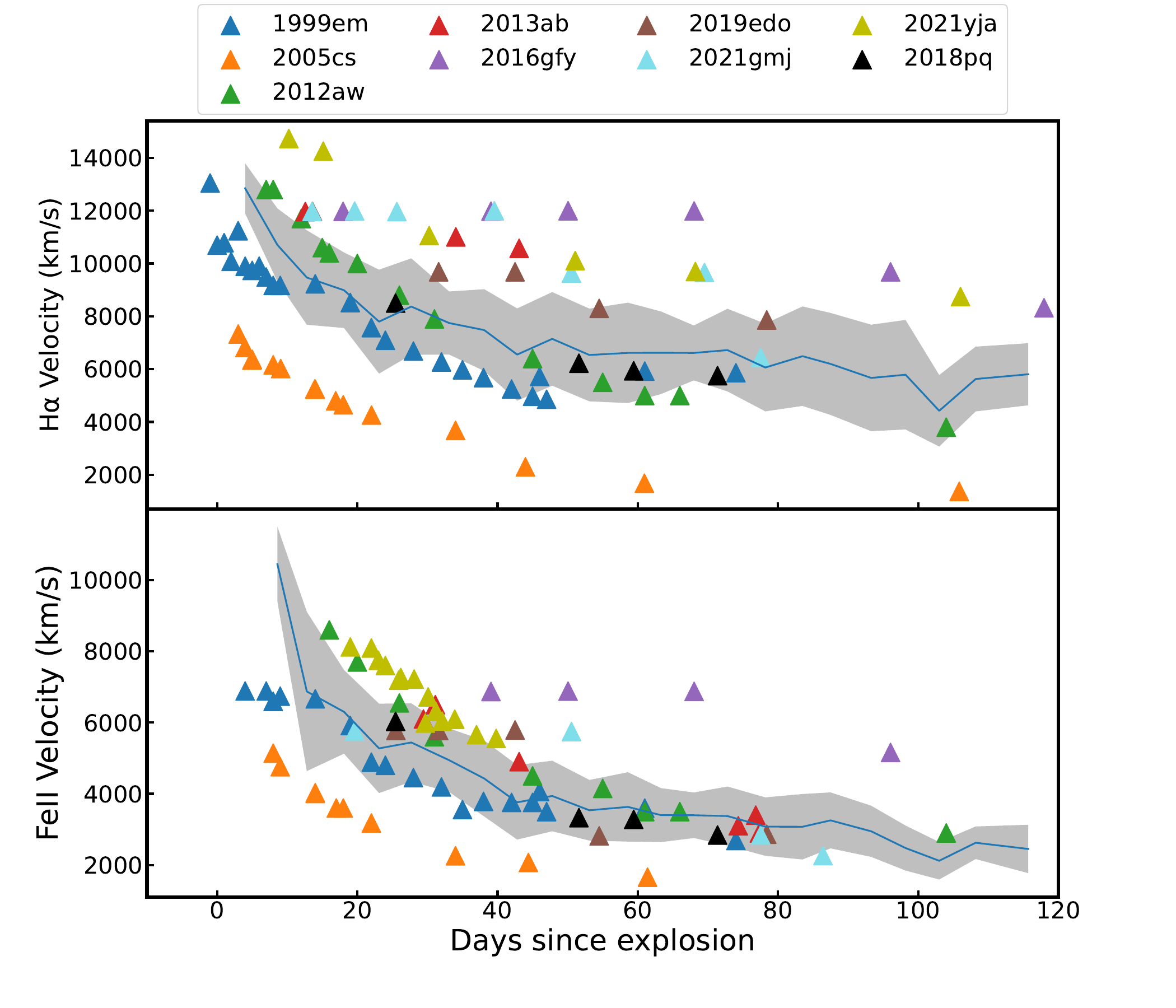}
    \caption{Upper Panel: Comparison of $H{\alpha}$ line velocity of SN~2018pq with other Type IIP SNe. Lower Panel: Comparison of Fe II (5169\AA) line velocity of SN~2018pq with other Type IIP SNe along with mean velocity (blue line) and standard deviation of mean velocities (gray shaded region) of 122 Type IIP/IIL sample \citep{Gutierrez_2017}.}
    \label{fig:Vel_evo}
\end{figure}

In the first spectrum of SN~2018pq, the P-Cygni profile of the $H\alpha$ (6563 \AA) line is visible on top of the continuum. The emission counterpart of $H\alpha$ profile is larger than the other spectra, indicating higher photospheric velocity at this epoch than the later epochs (see Figure~\ref{fig:Vel_evo}). Small peaks of metal lines are also visible in this spectrum as the photosphere recedes inward. In the mid-plateau spectrum of 51.6 d, metal lines are much clearer as they emerge from the photosphere. The emission line profile of $H\alpha$ becomes narrower with time, as the ejecta expands and becomes less dense, we see less number of photons scattered from high-velocity regions relative to regions closer to the photosphere, which is receding inward into the ejecta. The $H\beta$ (4861 \AA) and Na I D (5893 \AA) lines are visible. Fe II (4924 \AA, 5018 \AA, 5169 \AA), Ca II NIR triplet (8498 \AA, 8542 \AA, 8662 \AA), and Sc II (6245 \AA) lines are visible prominently at 51.6 d and later spectra. The absorption profile of Fe II and Ca II NIR lines becomes stronger with time. At 71.4 d, the SN enters a cooler photosphere phase when the photosphere penetrates the deeper layers rich in heavier metals. As the last spectrum (79.5 d) is noisy and difficult to find any spectral features, smoothing is applied to it with a 1D filter of window length 18 and a 2nd-order polynomial,  revealing the presence of $H\alpha$ and Ca II NIR lines.

\begin{figure}
\includegraphics[width=1.0\linewidth ]{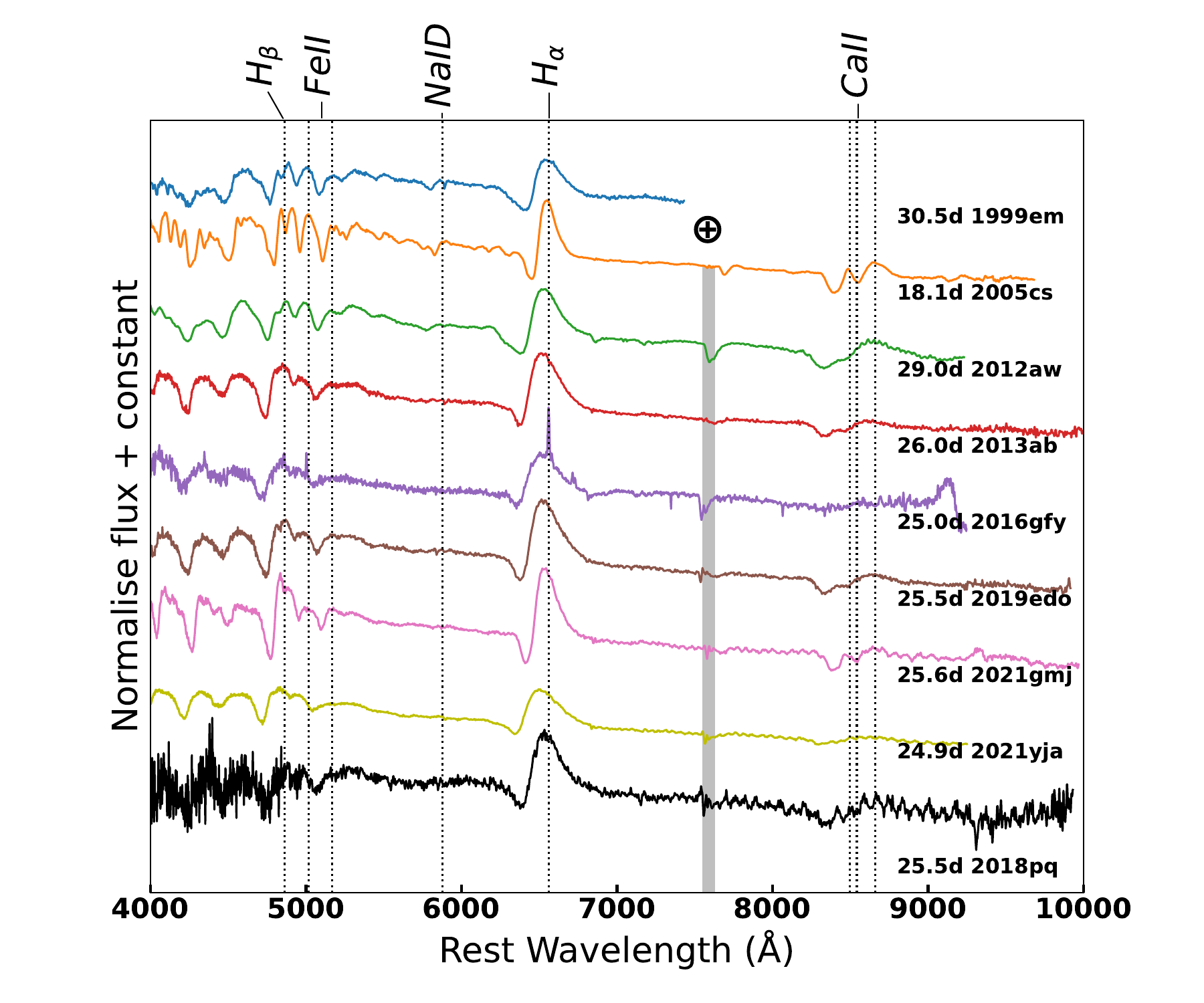} 
\caption{ Comparison of 25.5 d spectra of SN~2018pq with other Type IIP SNe during the early-plateau phase. The comparison sample is taken from Table~\ref{tab:comparison_objects}.}
\label{fig:Spec_15}
\end{figure}

\begin{figure}
\includegraphics[width=1.0\linewidth ]{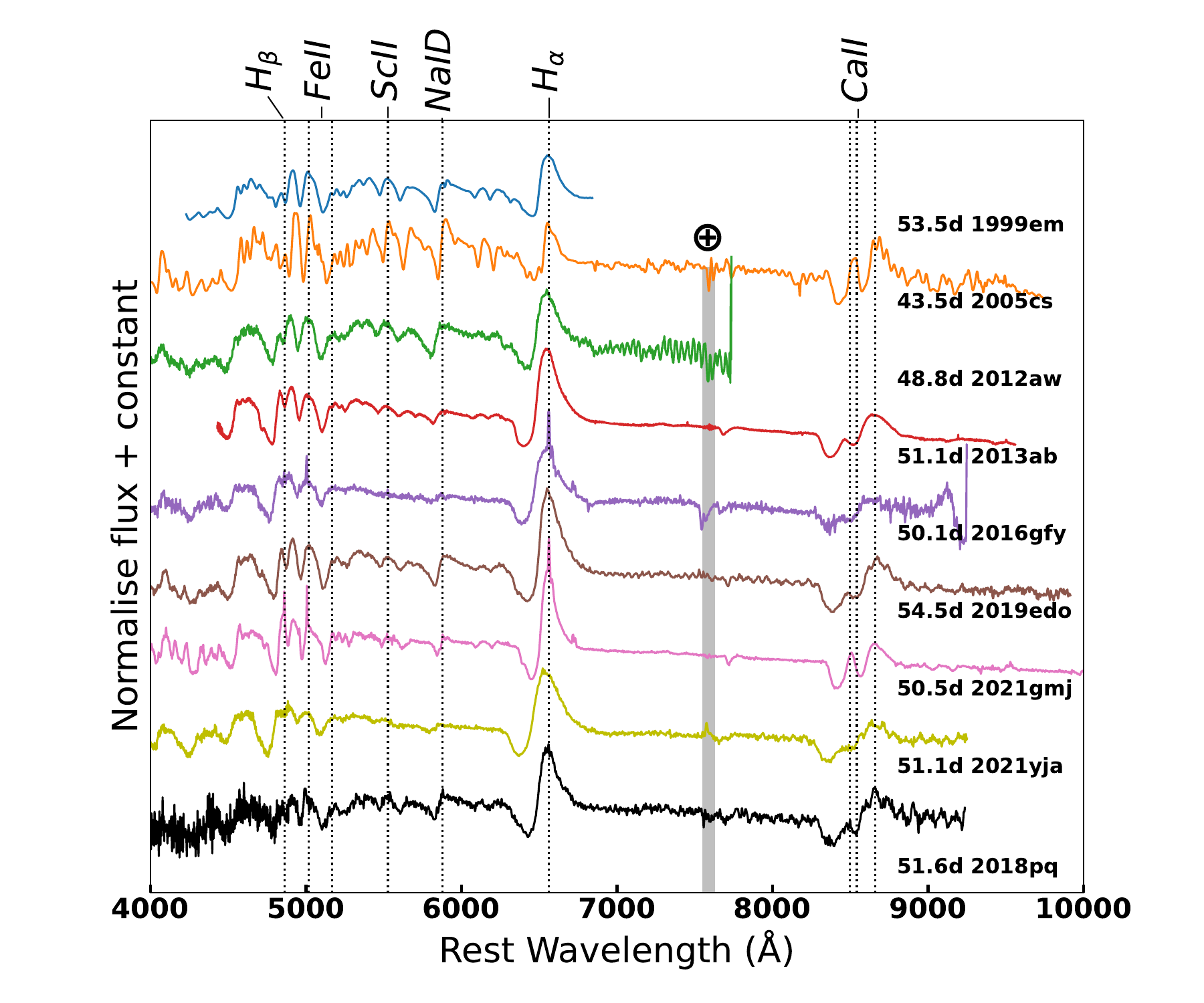} 
\caption{Comparison of the 51.6 d spectrum of SN~2018pq with other Type IIP SNe during the mid-plateau phase. The comparison sample is taken from Table~\ref{tab:comparison_objects}. }
\label{fig:Spec_41}
\end{figure}

In Figures~\ref{fig:Spec_15} and \ref{fig:Spec_41}, the spectra of SN~2018pq at two epochs (25.5 d and 51.6 d) are compared with Type IIP SNe of the comparison sample. The $H\alpha$ line is well matched with all SNe at both epochs, confirming SN~2018pq to be a Type IIP SN. Metal line profiles like Fe II and Ca II are also well-matched. SN~2018pq bears a close resemblance in terms of spectral line features with SNe~1999em, 2012aw, 2013ab, and 2019edo. Compared to other SNe, SN~2018pq exhibits a weaker $H\beta$ line, indicating a lower temperature in the ejecta. A narrow emission line (6563 \AA) is apparent on top of the continuum in the spectrum of SN~2016gfy, indicating the emission from its host H II region \citep{2016gfy_singh}. The contamination of the nearby H II region is also visible in the spectra of SN~2021gmj \citep{2021gmj_Murai}. The width of the $H\alpha$ P-Cygni profile in SN~2018pq is similar to SN~2021gmj in the 51.6 d spectra. This is wider than the low luminous SN~2005cs but narrower than normal SNe~2019edo, 2013ab, and long-plateau SN~2021yja.

The $H{\alpha}$ and Fe II (5169\AA) line velocities were calculated from the blue-shifted minima measured by fitting a Gaussian in each line profile in the spectra at four different epochs (25.5 d, 51.6 d, 59.4 d, and 71.4 d ) to reveal the velocity evolution. In Figure \ref{fig:Vel_evo}, $H{\alpha}$ and Fe II line velocities of SN~2018pq have been shown along with the SNe of the comparison sample and the mean velocity level of the 122 samples \citep{Gutierrez_2017}, illustrated by the shaded grey region. The rate of expansion of the outermost layer is represented by $H_{\alpha}$ line velocity, which is higher than photospheric velocity, represented by Fe II line velocity. The velocity evolution of SN~2018pq is lower than the mean velocity level, indicating slow-evolving SN, and similar to the velocity evolution of SN~1999em. The line velocities continuously decrease as the photosphere moves deeper inside the ejecta. The velocity evolution of SN~2018pq is more or less like a normal Type IIP SNe as it is higher than the sub-luminous SN~2005cs and lower than long-plateau SN~2021yja.

\subsection{\textsc{TARDIS} radiative transfer modelling}

We performed the spectra modelling of SN~2018pq using a version of the radiative transfer code \textsc{TARDIS} \citep{Kerzendorf2014}, which was repurposed for the early photospheric phase of Type II SNe \citep{Vogl2019}. In the code, H excitation and ionization are treated under non-local thermodynamic equilibrium (NLTE) conditions, which allows for the accurate modelling of the Balmer line series. The code treats the ejecta as spherically symmetric and homologously expanding, parametrised only by a handful of input parameters. The modelling is further simplified by the use of power-law density profiles (in the form of $\rho = \rho_0(r/r_0)^{-n}$, where $\rho_0$ denotes the density at a characteristic radius $r_0$, and $n$ representing the power-law steepness) and a uniform chemical composition, which are well-motivated assumptions in the photospheric phase of Type II SNe (see \citealt{DessartHillier2006, DessartHillier2008, Vogl2019}). However, the code does not account for time-dependent effects, and it also neglects the NLTE treatment of iron group elements, both of which become significant with the metal line blanketing a couple of weeks after the explosion. To this end, the standard \textsc{TARDIS} Type II SN modelling is most applicable in the first month of the SN evolution. 

To fit the \textsc{TARDIS} spectra to the SN~2018pq observations, we employed the spectral emulator of \cite{Vogl2020}. During the fitting, we followed the strategy described in \cite{Vasylyev2022, Vasylyev2023} and \cite{Csoernyei2023a, Csoernyei2023b}, performing the maximum likelihood-based fitting of the spectra. \cite{Csoernyei2023b} describes the most up-to-date grid of models used to train this emulator and the fitting. For the fitting, the spectra are corrected for the galactic reddening component of $E(B-V) = 0.186$ mag towards the SN as mentioned in Section~\ref{introduction} and normalised to the maximum flux. We then employed the emulator to infer the photospheric temperature ($T_\mathit{ph}$), photospheric velocity ($v_\mathit{ph}$), and the steepness of the density profile ($n$) for each spectral epoch. For the fitting, we have masked the O$_2$ band telluric regions.

\begin{figure}
    \centering
    \includegraphics[width=1.0\linewidth]{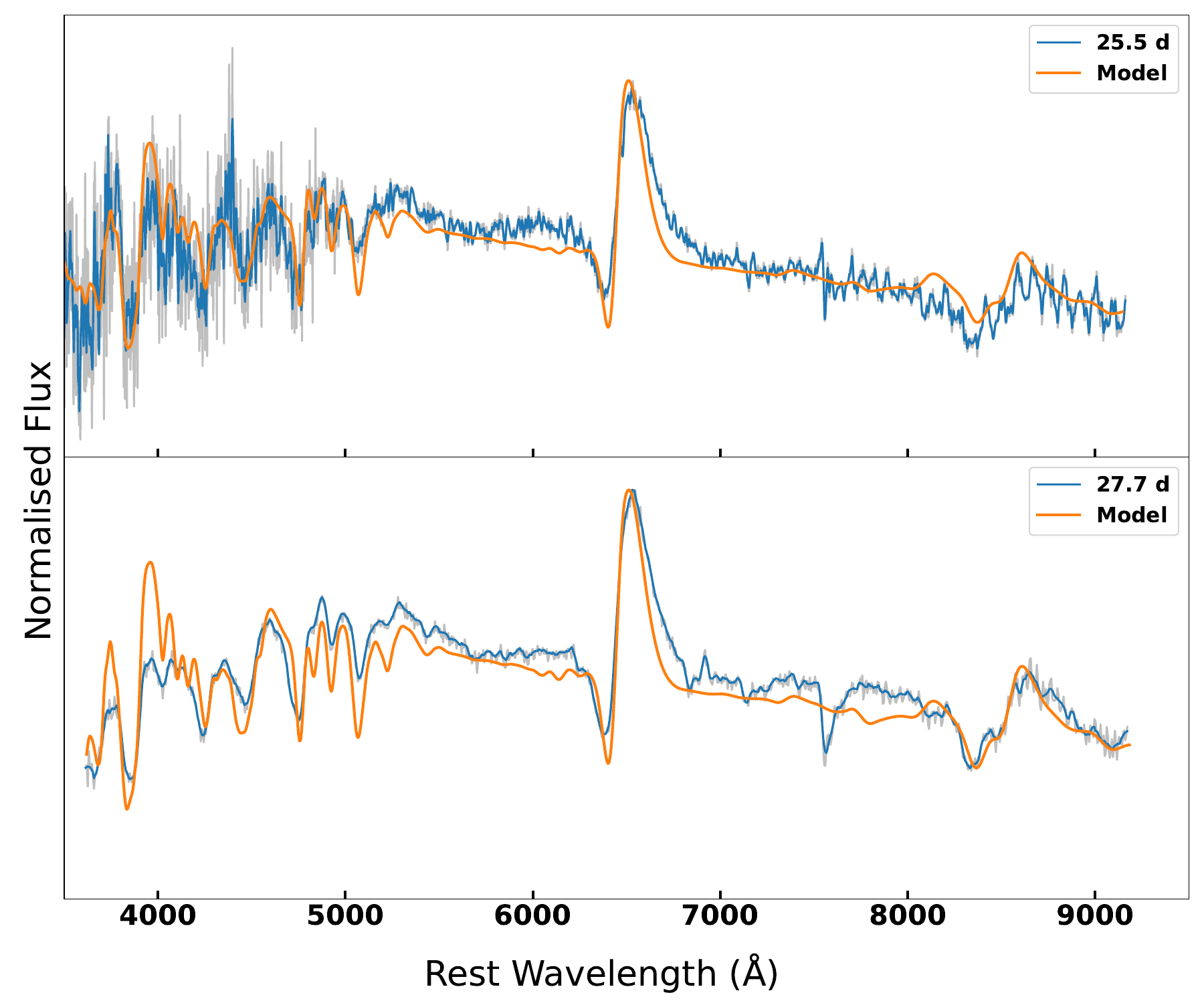}
    \caption{The best-fit spectra obtained from \textsc{TARDIS} modelling for the epochs 25.5 d, and 27.7 d of SN~2018pq.The unsmoothed and smoothed spectra are presented in gray and blue colour, respectively.}
    \label{fig:tardis_0.18}
\end{figure}

Figure~\ref{fig:tardis_0.18} shows the best-fit spectra for the 25.5 and 27.7 d of SN~2018pq. The fits show a good agreement with the spectra, with most of the metal line blanketing region, Balmer lines, and the Ca II triplet features reproduced adequately. The most significant qualitative mismatch is seen between the strength of lines corresponding to Fe II/III and Ca II in the blanketing region, which are predicted to be stronger by the \textsc{TARDIS} models, and are a result of the LTE treatment for the iron group elements as discussed above. Beyond this mismatch, the good fit indicates that SN~2018pq is a spectroscopically normal Type II SN. The best-fit parameters obtained from the modelling are displayed in Table~\ref{tab:tardis}. 

\begin{table}
    \centering
    \caption{Best fit physical parameters obtained through the \textsc{TARDIS} modelling.}
    \label{tab:tardis}
    \begin{tabular}{cccc}
        \hline
        Epoch & n & $v_{\textit{ph}}$ [km/s] & $T_{\textit{ph}}$ [K] \\
        \hline
         25.5 & 8.9 & 5662 & 6031 \\
         27.7 & 8.6 & 5642 & 5998 \\
        \hline
    \end{tabular} 
\end{table}

\section{Light curve modelling}
\label{sec:modelling}

\subsection{Semi-analytical modelling}
\label{sec:semi_analytical}

\begin{figure}
    \centering 
    \includegraphics[width=1.0\linewidth]{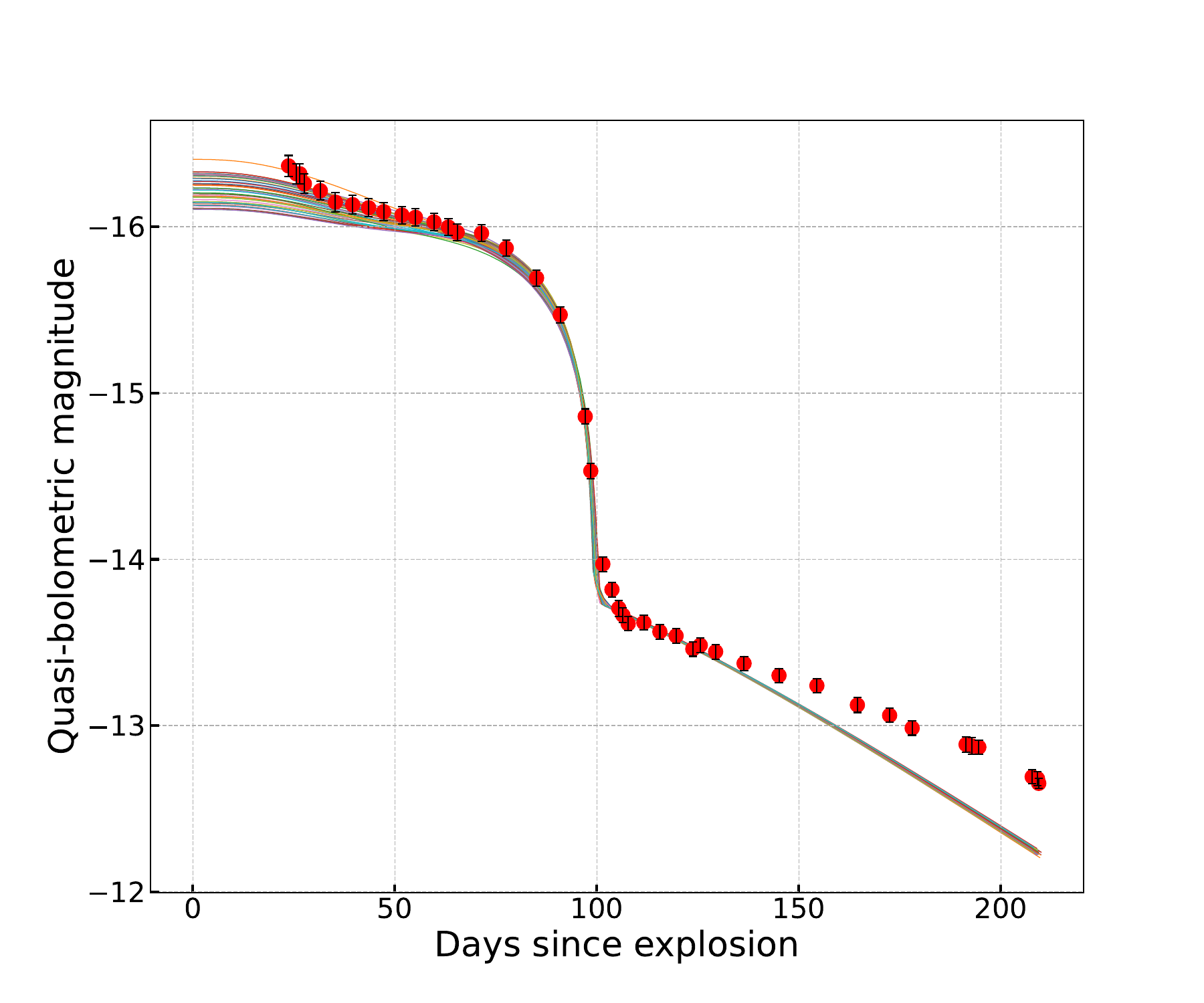}
    \caption{The quasi-bolometric light curve of SN~2018pq with the 50 best-fit light curves is shown, following \citet{Nagy_2014}, (\citeyear{2016A&A...589A..53N}) and \citet{10.1093/mnras/staa1743}.}
    \label{fig:n&v_Qbol}
\end{figure}

The quasi-bolometric light curve of SN~2018pq was made by using {\em BgVri} bands using \texttt{SuperBol} \citep{Nicholl_2018}, excluding the contributions from the infrared (IR) and ultraviolet (UV) regions. \cite{10.1093/mnras/staa1743} modified the \cite{Nagy_2014} model to obtain the best-fit core parameters and their uncertainties by including the MCMC method. This model assumes a spherically symmetric SN ejecta expanding homologously, with a core of uniform density, and outer layers with an exponential density profile. The model fits only the core part and hence, uses the light curve $\sim$30 d after the explosion. The radiation transport used by this code is based on the radiation diffusion model, provided by \cite{1989ApJ...340..396A}. For modelling, we sampled 4 main initial properties of the progenitor star in the input file: radius ($R_0$), ejecta mass ($M_\mathit{ej}$), kinetic energy ($E_k$), and thermal energy ($E_\mathit{th}$). \cite{1989ApJ...340..396A} found a correlation between the parameters using the Pearson correlation coefficient method, revealing that the two pairs of parameters: $M_\mathit{ej}$ -- $E_k$ and $E_\mathit{th}$ -- $R_0$, are highly correlated. The recombination temperature and Thomson scattering opacity ($\kappa$) were kept at 5500~K and 0.3~cm$^2$/g, respectively, while performing the modelling. In Figure~\ref{fig:n&v_Qbol}, 50 best-fit light curves to the quasi-bolometric light curve of SN~2018pq are shown. The plateau and drop from the plateau to nebular phase are well reproduced by the model. The best fit model estimates the $M_\mathit{ej}$ = 10.7$^{+0.3}_{-1.8}$ M$_\odot$, with an $E_k$ =  2.6$^{+0.2}_{-0.7}$$\times$10$^{51}$ ergs. Assuming the proto-neutron star mass of 2 M$_\odot$, the total progenitor mass is inferred to be around 13 M$_\odot$. The plateau duration is highly affected by the $M_\mathit{ej}$, $R_0$, and $E_k$, whereas the tail part is highly dependent on the $^{56}$Ni mass and gamma-ray leakage. Here, the effect of magnetar was not considered. The recombination temperature has a minimal effect on the light curve. The modelling was performed on the quasi-bolometric light curve; therefore, the estimated explosion parameters should be considered as the lower limits. The model does not reproduce the tail part of the quasi-bolometric light curve beyond 140 days, as, at this stage, the SN ejecta become optically thin, and the tail luminosity is governed by several factors such as asymmetries and the mixing of nickel into the outer layers. These factors are not considered in the model. Therefore, the model could not fit the observed light curve during late phases, and the obtained $^{56}$Ni mass from the model is not reliable. The values of corresponding physical parameters with their 1$\sigma$ uncertainties and the prior range are provided in Table~\ref{tab:N & V results}. 

\begin{table}
	\begin{center}
	\caption{The best-fit core parameters with 1$\sigma$ uncertainties for the quasi bolometric light curve of SN~2018pq, following \citet{ Nagy_2014}, (\citeyear{ 2016A&A...589A..53N}) and \citet{10.1093/mnras/staa1743}. }
        \label{tab:N & V results}
	\begin{tabular}{llll} 
		\hline
		Parameter & Best-fit value & Prior range\\
            \hline
            \vspace{0.3cm}
            Initial radius [$R_0$($10^{13}$cm)] & $2.95^{+0.05}_{-1.78}$ & 0.5--3 \\
            \vspace{0.2cm}
             Ejecta mass [$M_\mathit{ej}$ ($M_\odot$)] & $10.7^{+0.3}_{-1.8}$ & 7--15\\
            \vspace{0.2cm}
            Kinetic energy [$E_k$ (10$^{51}$ ergs)] & $2.6^{+0.2}_{-0.7}$ & 0.1--3 \\
            \vspace{0.2cm}
            Thermal energy [$E_\mathit{th}$ (10$^{51}$ ergs)] & $0.48^{+0.71}_{-0.04}$ & 0.01--3\\
            \hline
		
	\end{tabular}
        \end{center}
\end{table}

\subsection{Hydrodynamical modelling: \textsc{MESA+STELLA}}
\label{sec:hydrodynamical}
 
We employ 1D hydrodynamical modelling using the Module for Experiments in Stellar Astrophysics \citep[\textsc{MESA};][]{2015ApJS..220...15P}, version r-15140, combined with the radiative transfer code, \textsc{STELLA} \citep{1998ApJ...496..454B} for modelling the quasi-bolometric light curve and exploring the explosion and progenitor properties of SN~2018pq. A detailed description to simulate a CCSN from a zero-age main sequence star (ZAMS) is given in \cite{2018ApJS..234...34P}. Inside a star, convection is one of the important phenomena through which energy is transported. The convective boundary is a challenging region where the radial velocity of the bulk motion of the matter goes to zero. It is important to place the convective boundaries properly, as their location strongly influences stellar evolution. During simulation, convective boundaries are well placed by \textsc{MESA}, fulfilling both the Schwarzschild \citep{Schwarzschild_1958} and the Ledoux \citep{Ledoux_1947} criteria.

\textsc{MESA} has incorporated element diffusion, a process by which particles are transported inside a star due to the gradients in density, temperature, and pressure, which are important for nuclear reaction and stellar structure. After the core-collapse, the outgoing shock waves cause Rayleigh-Taylor instabilities (RTI) in the outer envelope, and the matter is mixed beyond the shock front (\citealt{1976ApJ...207..872C}; \citealt{1978ApJ...219..994C}). Using shock-capturing hydrodynamics, \textsc{MESA} can model RTI in 1D \citep{2016ApJ...821...76D}. The afterwards shock breakout and evolution are handled by a frequency-dependent radiation hydrodynamic code, \textsc{STELLA}. It solves the conservation equations (mass, momentum, and energy) in a Lagrangian co-moving frame and the radiative transfer equations with the intensity momentum approximation in each frequency bin. An artificial viscosity based on the standard Von Neumann artificial viscous pressure \citep{1950JAP....21..232V} is also added. \textsc{STELLA} creates an opacity table built on over 153,000 spectral lines taken from \cite{1995all..book.....K} and \cite{1996ADNDT..64....1V}. The expansion velocity formalism from \cite{1993ApJ...412..731E} is used for line opacities, taking high-velocity gradients into account. Photoionisation, free-free absorption, and electron scattering are included in opacity. Plasma is assumed to be in LTE, where the Boltzmann-Saha distribution law is used for ionisation and level populations. Detailed nucleosynthesis reactions are not included in \textsc{STELLA} except the radioactive decay chain starting from $^{56}$Ni. The energy from the radioactive decay process is deposited through positrons and gamma-rays \citep{1995ApJ...446..766S}. The CSM interaction with SN ejecta can also be incorporated into the model to better replicate the early-time light curve, if required. As an output, \textsc{STELLA} produces bolometric and quasi-bolometric light curves. It also gives Fe II line velocities at different optical depths ($\tau$ = 0.2, 1.0, 2.0) along with the properties of the photosphere (such as radius, temperature, mass, opacity), broadband light curves in {\em U}, {\em B}, {\em V}, {\em R} and {\em I} bands, gamma ray deposition, and a rough estimate of the effective temperature.

Different test suites dedicated to CCSN modelling are used to produce the light curves. At first, \texttt{make$\_$pre$\_$ccsn} is used to evolve a star from pre-main-sequence to Fe core infall by changing the different initial parameters of the star (such as initial mass, $M_\mathit{ZAMS}$; metallicity, $Z$; rotation, $(v/vc )_\mathit{ZAMS}$; wind scaling factor, $\eta_\mathit{wind}$; the mixing length in the H envelope, $\alpha_\mathit{MLT, H}$; etc). In the second step, the output is inserted into another test suite named \texttt{ccsn$\_$IIp}, which evolves the star till near shock breakout. In this module, explosion energy ($E_\mathit{exp}$) and $^{56}$Ni mass can be used as an input to modify the light curves. In the third step, the outcomes are handed off to \textsc{STELLA}, which handles shock-breakout and post-explosion evolution. The effect of some parameters is more significant than others on the synthetic light curves, like $E_\mathit{exp}$, $M_\mathit{ZAMS}$, and $^{56}$Ni mass. In contrast, the effect of $Z$ and rotation are insignificant. While comparing the synthetic light curve with the observed one, degeneracy in progenitor mass is noticed. This degeneracy can be removed by comparing the observed and synthetic Fe II line (5169 \AA) velocities, measured at Sobolev optical depth, $\tau_\mathit{sob}$ = 1.

\begin{figure}
    \centering
    \includegraphics[width=1.0\linewidth]{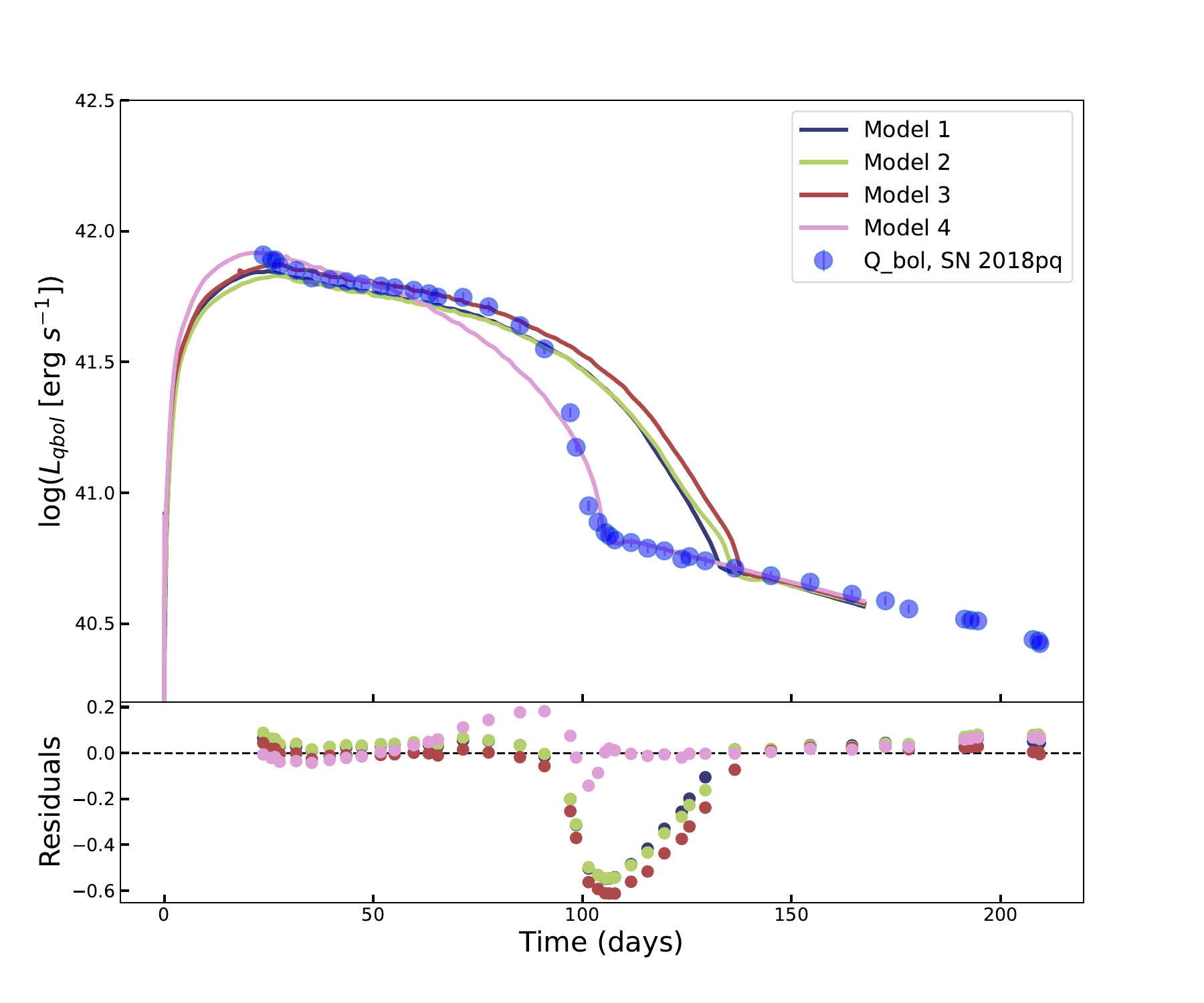}
    \caption{Synthetic light curves fitting generated using \textsc{MESA+STELLA} on the quasi-bolometric light curve of SN~2018pq. The obtained parameters from the models are summarised in Table~\ref{tab:MESA_results}.}
    \label{fig:mesa_qbol}
\end{figure}

In Figure~\ref{fig:mesa_qbol}, the best-fitting model light curves to the quasi-bolometric light curve of SN~2018pq are shown. The models are generated using solar metallicity ($Z$ = 0.02 $M_\odot$), and the default value of $\alpha_\mathit{MLT, H} = 3$. From semi-analytical modelling, we estimated the initial mass of the progenitor to be around 13 $M_\odot$ (see Section~\ref{sec:semi_analytical}). With \textsc{MESA+STELLA}, we generated light curve models starting with an initial progenitor mass of 13 $M_\odot$, ranging up to 16 $M_\odot$. As the early light curve ($\leq$ 23 d) is not available CSM interaction is not included in the models. The lowest mass model of 13 $M_\odot$ overestimated the plateau length. We increased the wind scaling factor and rotation to fit the plateau and the transition between plateau and nebular phase, but the 13 $M_\odot$ model does not explode when the wind scaling factor changes from 1 to 2 and rotation from 0.0 to 0.4. The model light curve from a 14 $M_\odot$ progenitor (model 1) fits the plateau relatively well; however, the transition phase from the plateau end and the radioactive tail is overestimated. The initial mass is increased to 16 $M_\odot$ unless the model fits the transition region. Models 1, 2, and 3 with initial masses of 14, 15, and 16 M$_\odot$, respectively, and wind scaling factor 1 reproduce the plateau phase well, while model 4 with an initial mass of 16 M$_\odot$ and wind scaling factor 2 replicates the transition phase between the plateau and radioactive tail phase. All models are well-fitted at the radioactive tail phase. The variations in explosion energy and $^{56}$Ni in these models are very small. With increasing initial mass and constant wind scaling factor, the radius and final mass of the progenitor increase. If the wind scaling factor increases from 1 to 2, the radius increases slightly, but the final mass decreases, as seen for models 3 and 4. This signifies that mass loss is enhanced due to the strengthening of stellar wind. As seen in Figure~\ref{fig:mesa_qbol}, none of the model light curves reproduce the entire observed light curve. Sometimes, there are difficulties noticed in reproducing the entire light curve with a single model \citep{Hiramatsu_2021, Teja_2022, Forde_2025}. The models 1, 2 and 3 exhibit less stellar wind; hence, the progenitor has a larger H envelope, resulting in a longer plateau duration phase. Although these models replicate the slow decline of the plateau, they fail to recreate the sharp transition between the plateau and the tail phase. The enhanced RTI mixing can produce this type of sharp decline \citep{2018ApJS..234...34P}, hence, at the end of the plateau phase, the mixing of the SN-ejecta is underestimated by the models. Model 4 exhibits more stellar wind, leading to a lower amount of H to be left on the progenitor before the explosion. This model reproduced the sharp decline during the transition but overestimated the slope of the plateau. The progenitor may not shed as much envelope mass through stellar wind prior to the explosion as the model considers. Therefore, the progenitor may have had a well-mixed, high envelope mass before the explosion compared to the other models considered.

\begin{figure}
    \centering
    \includegraphics[width=1.0\linewidth]{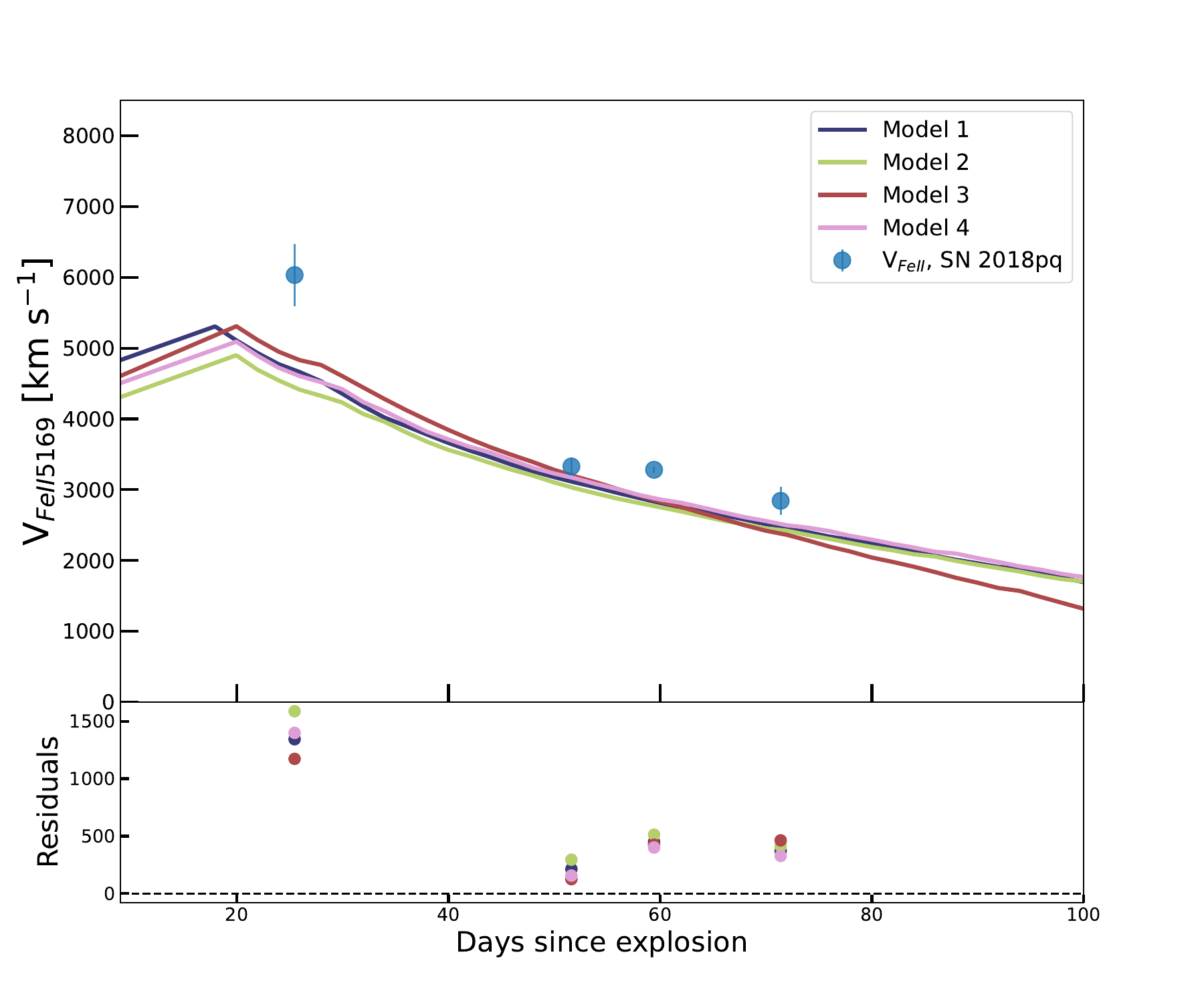}
    \caption{The evolution of photospheric velocities obtained
from \textsc{MESA+STELLA} modelling for the different models with an optical depth ($\tau_\mathit{sob}$) = 1.0 compared with the observed photospheric velocities.}
    \label{fig:mesa_vel}
\end{figure}

To remove degeneracy in mass, the photospheric velocity (Fe II 5169 \AA) obtained from all models is compared with the observed velocities in Figure~\ref{fig:mesa_vel}. The model velocities do not show any significant difference. In conclusion, hydrodynamical modelling suggests a progenitor in the mass range of 14 to 16 $M_\odot$, which is higher than the results obtained from the analytical modelling. The different models and their corresponding estimates are summarised in Table~\ref{tab:MESA_results}. 

\begin{table*}
    \centering
    \caption{Initial parameters of the progenitor obtained from different models generated in \textsc{MESA+STELLA}}
    \label{tab:MESA_results}
    \begin{tabular}{|cccccccc|}
        \hline
        Model &  $M_{ZAMS}$ ($M_\odot$) & $E_\mathit{exp}$ (10$^{51}$ ergs) & $M_\mathit{Ni} (M_\odot$) & $(v/v_c)_\mathit{ZAMS}$ & $\eta_\mathit{wind}$ & R ($R_\odot$) & $M_f$ ($M_\odot$)\\
        \hline
        1  & 14 & 0.35 & 0.024 & 0.50 & 1 & 646.2 & 11.9  \\
        2  & 15 & 0.30 & 0.020 & 0.50 & 1 & 718.4 & 12.4  \\
        3  & 16 & 0.35 & 0.022 & 0.30 & 1 & 743.5 & 13.2  \\
        4  & 16 & 0.30 & 0.020 & 0.40 & 2 & 771.2 & 10.2  \\
        \hline
    \end{tabular}
   
\end{table*}

\section{Discussion and conclusion}
\label{sec:discussion}

\begin{figure}
    \centering
    \includegraphics[width=1.0\linewidth]{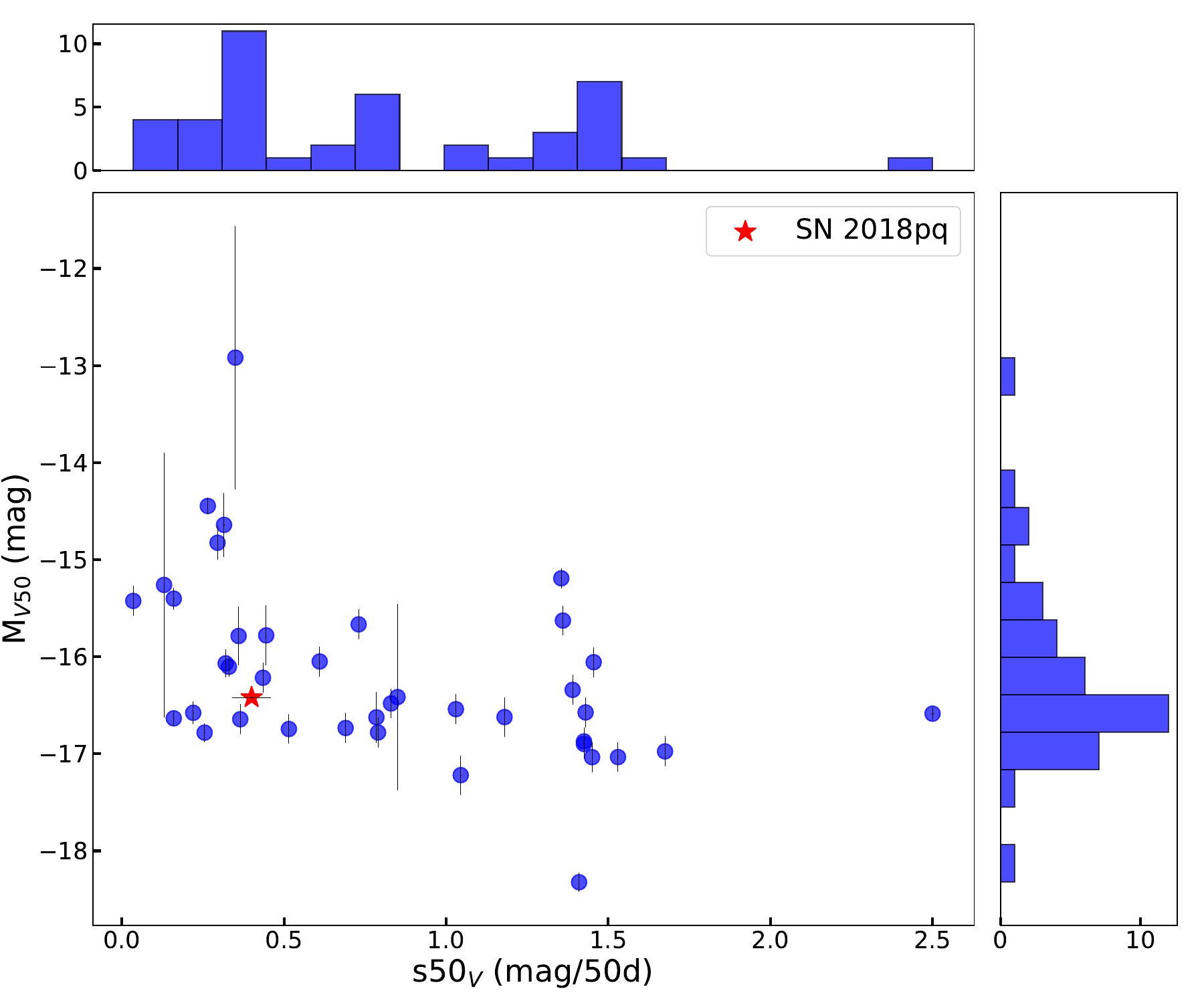}
    \caption{The correlation between the {\em V}-band magnitude at 50 d post-explosion (M$_\mathit{V50}$) and slope of the {\em V}-band light curve from peak magnitude to 50 d since explosion (s50$_V$).}
    \label{fig:mv50_sv50}
\end{figure}

\begin{figure}
    \centering
    \includegraphics[width=1.0\linewidth]{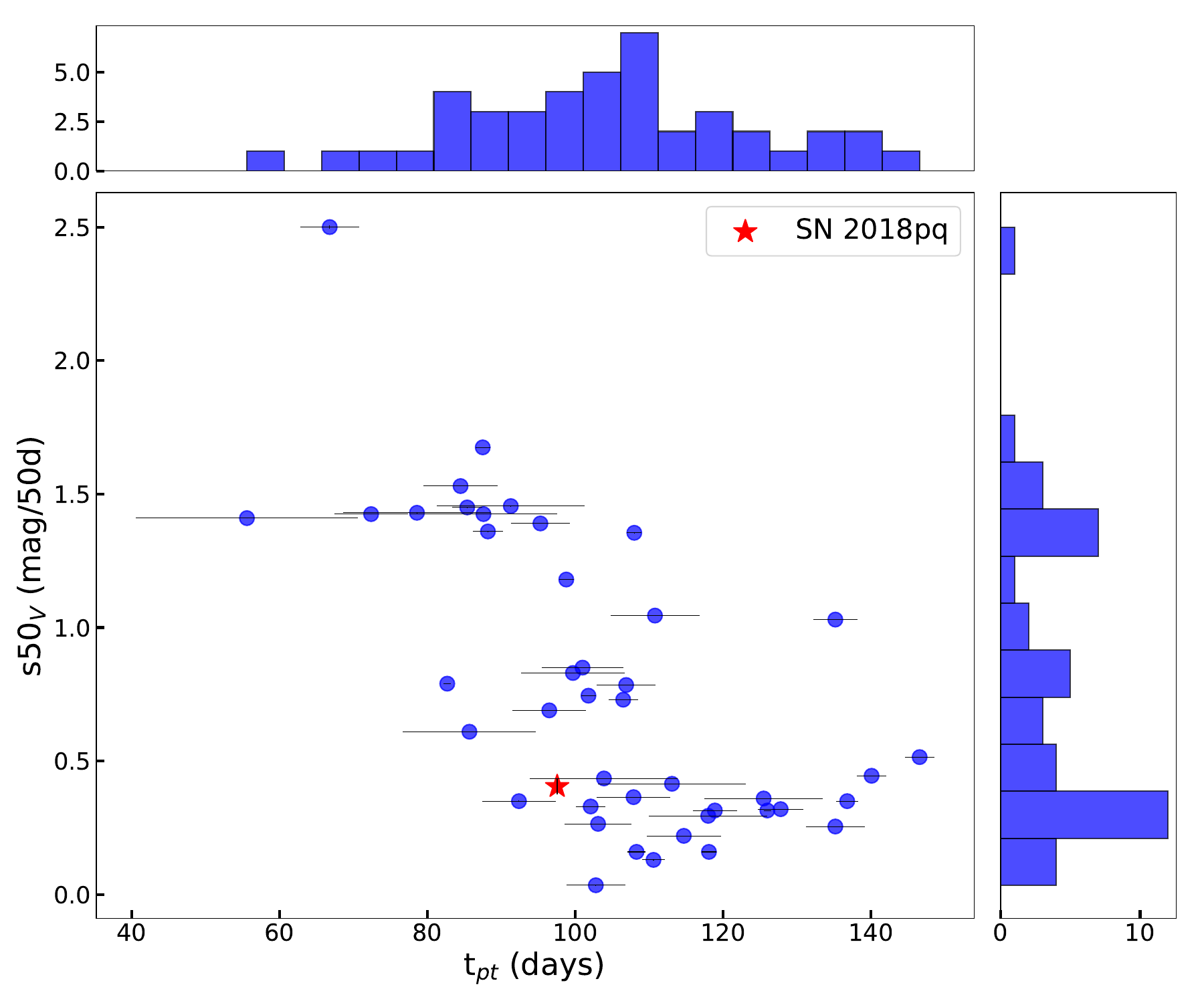}
    \caption{The correlation between the plateau duration (t$_\mathit{pt}$) and slope of the {\em V}-band light curve from peak magnitude to 50 d since explosion (s50$_V$).}
    \label{fig:s50v_tpt}
\end{figure}

The photometric and spectroscopic characteristics of SN~2018pq classify it as a Type IIP SN. The first photometric detection of SN~2018pq was approximately 23 d after the estimated explosion epoch (JD 2458135.45$\pm$3.48) when the SN had already entered the plateau phase. The spectral evolution reveals no NaID absorption line, which aligns with the position of the SN at the outskirts of the galaxy. Therefore, only the line-of-sight galactic extinction ( $E(B-V) = 0.186$ mag) was considered in the analysis, as no additional host galaxy extinction was inferred. The absolute {\em V}-band magnitude measured at 50 d post-explosion was estimated to be $-$16.42 $\pm$ 0.01 mag, with a plateau length of $\sim$97 d.

The observational parameters of SN~2018pq show notable similarities to the typical Type IIP SN~1999em. The plateau phase of SN~2018pq persists for $\sim$97 d, similar to SNe~1999em (95 d) and 2012aw (96 d). It falls into the category of normal luminous Type IIP events with a decay rate (s50$_V$) of $0.42\pm0.06$ mag 50 d$^{-1}$ during the plateau phase, comparable to SN~2013ab (s50$_V$ = $0.46$ mag 50 d$^{-1}$ ). The correlation plots in Figures~\ref{fig:mv50_sv50} and \ref{fig:s50v_tpt} compare s50$_V$ with the magnitude at 50 d post explosion in {\em V}-band ($M_{V50}$) and the plateau duration ($t_\mathit{PT}$). In these plots, the Type II SNe sample is taken from the literature sample study \citep{Anderson_2014, Valenti_2016}.  In Figure~\ref{fig:mv50_sv50}, the parameter trend shows that low luminous SNe decline slowly, whereas the more luminous SNe decline faster during the plateau phase. SN~2018pq has a normal luminosity and follows an average decline rate (s50$_V$ = $0.42\pm0.06$ mag 50 d$^{-1}$). From Figure~\ref{fig:s50v_tpt}, it is revealed that the SNe with a longer plateau duration tend to have lower decline rates. SN~2018pq is located at the lower side of the distribution, suggesting that the luminosity during the plateau phase did not change much. The other events with similar plateau duration show higher values of the slope as compared to SN~2018pq. One of the possible reasons could be the relatively slower velocity of the SN ejecta, resulting in constant luminosity for a longer duration. The magnitude drop (2.24$\pm$0.01 mag) during the transition between the plateau and nebular phases, inferred from the analytical fit to the {\em V}-band light curve, was found to be higher than the plateau drop seen in SN~1999em. This drop is inversely proportional to the luminosity coming from synthesised $^{56}$Ni mass during the explosion, which is $0.029\pm0.003$ $M_\odot$ for SN~2018pq, lower than the $^{56}$Ni mass found in SN~1999em but similar to SN~2016gfy. SN~2018pq exhibits a steeper decline rate of 11.87$\pm$1.68 mag 100 d$^{-1}$ during the transition phase compared to other typical Type IIP SNe, a trend also observed in SN~2013ab, which showed a decline of 8.7$\pm$0.2 mag 100 d$^{-1}$. In the nebular phase, the decay rate of 0.99$\pm$0.03 mag 100 d$^{-1}$ aligns well with the decay rate expected from $^{56}$Co to $^{56}$Fe. 

The spectroscopic observation of SN~2018pq started two days after the photometric detection. The presence and evolution of the prominent P-Cygni profile of $H\alpha$ indicates the progenitor has an H envelope before the explosion. Metal lines become apparent in the later spectra, reflecting the cooling and recombination processes as the SN evolves. The spectral features show striking similarities with other Type IIP SNe considered in the sample. The observed $H\alpha$ and Fe II line velocities evolve like those in SNe~1999em and 2012aw. We performed the radiative-transfer modelling of two epochs of SN~2018pq using \textsc{TARDIS}. The velocity, photospheric temperature, and steepness of the density profile are obtained for each epoch from the best-fit modelling using only Galactic reddening. These results provide insights into the early expansion dynamics and thermal properties of the SN. By accurately modelling the Balmer series under NLTE consideration of H excitation and ionisation, the model reproduces the Balmer lines and the Ca II triplet with a good match to the observed spectra.

Semi-analytical and hydrodynamical modelling on the quasi-bolometric light curve of SN~2018pq has been carried out to estimate the physical parameters and explosion characteristics of the progenitor. The ejecta mass of around 11 $M_\odot$ estimated from the semi-analytical modelling is similar to the ejecta mass found in SN~1999em. However, the estimated radius and energy are higher than those of the sample SNe. \textsc{MESA+STELLA} hydrodynamical modelling suggests a higher progenitor mass in the range of 14--16 $M_\odot$, similar to SN~2012aw, but with a higher radius between 640--772 $R_\odot$ and low explosion energy between 0.30--0.35$\times$10$^{51}$ ergs. The discrepancy between the results obtained from semi-analytical and hydrodynamical models has been noticed in many studies \citep{Chatzopoulos_2012, Szalai_2019}. This is caused by the simplified assumptions in semi-analytical modelling, such as a two-component model with a constant density inner core and an envelope with exponentially decreased density, grey opacity treatment, and independence of metallicity. On the other hand, \texttt{MESA+STELLA} accounts for complex physical processes such as calculating the position of convective boundaries, including element diffusion, and energy conservation during massive star explosions. It also incorporates a 1D capability for modelling effects like making a realistic opacity table, accounting for the effect of RTI, and also includes the radioactive decay chain of $^{56}$Ni. The progenitor mass obtained from the hydrodynamical code sometimes gives a higher mass than simple semi-analytical modelling (e.g. SN~2017eaw; \cite{Szalai_2019}), whereas the explosion energy derived from semi-analytical modelling sometimes gives a higher value (e.g. SN~2020jfo; \cite{Teja_2022}). The simplified assumptions of semi-analytical modelling overlook the details and complex physical phenomena, resulting in differences in the outcome as compared to the hydrodynamical modelling.

The overall analysis of SN~2018pq suggests it to be a Type IIP SNe with normal luminosity. It shows many photometric and spectroscopic similarities with SNe~1999em and 2012aw. With good coverage of photometric data spanning from the plateau to the nebular phase, well-constrained $^{56}$Ni mass, and detailed analysis of the progenitor properties make it a good sample to populate the normal Type IIP events.

\section*{Acknowledgements}
The authors thank the referee for providing constructive comments on the manuscript, improving the clarity of presentation. This work uses data from the Las Cumbres Observatory global telescope network. The LCO group is supported by NSF grants AST-1911151 and AST-1911225. C. Pellegrino acknowledges support from ADAP program grant No. 80NSSC24K0180 and NSF grant AST-2206657. MD acknowledges the Innovation in Science Pursuit for Inspired Research (INSPIRE) fellowship award (DST/INSPIRE Fellowship/2020/IF200251) for this work. KM, ND and BA acknowledge the support from the BRICS grant DST/ICD/BRICS/Call-5/CoNMuTraMO/2023 (G) funded by the Department of Science and Technology (DST), India. RD acknowledges funds by ANID grant FONDECYT Postdoctorado Nº 3220449.

\section*{Data Availability}
The data presented in this paper will be provided upon request. The spectra will be made publicly available in WiseRep and Zenodo.



\bibliographystyle{mnras}
\bibliography{SN2018pq} 

\bsp	
\label{lastpage}
\end{document}